\documentclass[aps,prl,twocolumn,amssymb,superscriptaddress]{revtex4}
\usepackage{subfigure}
\usepackage{graphicx}
\usepackage{rotating} 
\usepackage{color}
\usepackage{lscape}
\usepackage{rotating}
\usepackage{xfrac}
\usepackage{amsmath,amsthm}
\usepackage{epstopdf}
\usepackage{xr}

\begin{document}
\title{Ultraslow water-mediated transmembrane interactions regulate the activation of A$_{\text{2A}}$ adenosine receptor}

\author{Yoonji Lee}
	\affiliation{National Leading Research Laboratory (NLRL) of Molecular Modeling and Drug Design, College of Pharmacy and Graduate School of Pharmaceutical Sciences, Ewha Womans University, Seoul 03760, Korea}
	\author{Songmi Kim}
	\affiliation{Korea Institute for Advanced Study, Seoul 02455, Republic of Korea}
	\author{Sun Choi}
	\thanks{sunchoi@ewha.ac.kr}
	\affiliation{National Leading Research Laboratory (NLRL) of Molecular Modeling and Drug Design, College of Pharmacy and Graduate School of Pharmaceutical Sciences, Ewha Womans University, Seoul 03760, Korea}
	\author{Changbong Hyeon}
    \thanks{hyeoncb@kias.re.kr}
	\affiliation{Korea Institute for Advanced Study, Seoul 02455, Republic of Korea}

\date{\today}

\begin{abstract}
Water molecules inside G-protein coupled receptor have recently been spotlighted in a series of crystal structures. 
To decipher the dynamics and functional roles of internal waters in GPCR activity, 
we studied A$_{\text{2A}}$ adenosine receptor using $\mu$sec-molecular dynamics simulations. 
Our study finds that the amount of water flux across the transmembrane (TM) domain varies depending on the receptor state, and that the water molecules of the TM channel in the active state flow three times slower than those in the inactive state. 
Depending on the location in solvent-protein interface as well as the receptor state, the average residence time of water in each residue varies from $\sim\mathcal{O}(10^2)$ psec to $\sim\mathcal{O}(10^2)$ nsec. 
Especially, water molecules, exhibiting ultraslow relaxation ($\sim\mathcal{O}(10^2)$ nsec) in the active state, 
are found around the microswitch residues that are considered activity hotspots for GPCR function. 
A continuous allosteric network spanning the TM domain, arising from water-mediated contacts, is unique in the active state, underscoring the importance of slow waters in the GPCR activation.
\end{abstract}
\maketitle


\section*{INTRODUCTION}

The activation of GPCRs is associated with a conformational change of the receptor modulated by an agonist binding.
As suggested by the X-ray crystal structure of $\beta_2$-adrenergic receptor, the outward tilting of TM5 and TM6 helices that accompanies a helix rotation of the cytoplasmic part of TM5-ICL3-TM6 is one of the key structural changes that facilitate G-protein recruitment \citep{Rosenbaum2009Nature,rasmussen2011Nature,lebon2011Nature}.
More microscopically, orchestrated dynamics among a common set of highly conserved fingerprint residues, dubbed as \emph{microswitches}, of the class A GPCR family is critical for the regulation of GPCR activity \citep{nygaard2009TPS, katritch2013ARPT}.
These microswitches, including three ubiquitous sequence motifs, DRY, CWxP, and NPxxY (where `x' denotes any amino acid residue), adopt distinct rotameric configurations in their side chains, which are sensitive to the type of a cognate ligand bound to the orthosteric site \citep{Lee2015PLoSComp}.
In the agonist-bound state, the dynamic correlations among microswitches allow the receptor to adopt and maintain the active conformation, which facilitate accommodation of G-protein \citep{Lee2015PLoSComp}.

Besides the allosteric network defined by the inter-residue contacts \citep{Lee2015PLoSComp,Lee14Proteins}, water molecules can also contribute to the formation of the intra-receptor signaling network.
Given that GPCRs are the key therapeutic targets that sense and process external stimuli with an exquisite precision, it is critical to understand the roles played by water as well as the residue dynamics in GPCR activation.
As is well appreciated by many studies, 
water is essential for structure, dynamics, and function of biomolecules \citep{Tarek00BJ,Ball2008ChemRev,Frauenfelder2009PNAS}.
Water molecules that form hydration shell on protein surfaces \citep{Tarek00BJ,Tsai00BJ,Pal2002PNAS,Jungwirth2015JPCL} 
or a lipid-mediated effect of water via hydrophobic matching to the protein surface \citep{Mallikarjunaiah2011BJ} have long been studied.
Especially, possible functional roles of internal water molecules found inside TM domain in GPCR activation have recently been discussed \citep{Yuan2014NatCommun,Yuan2013Angewandte,Sun2014JPCB,burg2015Science,Leioatts2014Biochemistry}.
Recently, MD simulations studies have analyzed water in various type of GPCRs, including A$_{\text{2A}}$AR \citep{Yuan2014NatCommun,Lee2012BJ,Yuan2015Angewandte,Sabbadin2014JCIM}, $\beta_2$-adrenergic receptor \citep{bai2014PCCP}, rhodopsin \citep{Sun2014JPCB,grossfield2008JMB,Jardon2010BJ,Leioatts2014Biochemistry}, dopaminergic receptor \citep{Selent2010PLoSCompBiol}, and $\mu$-opioid receptor \citep{Yuan2013Angewandte}; yet the foci of these studies were mostly on mapping the locations of internal waters and metal ions inside TM domain, or on the local dynamics of water specific to the ligand binding cleft in the process of ligand-receptor recognition \citep{Sabbadin2014JCIM}.
\emph{Dynamical} properties of internal water over the entire architecture of GPCR have not fully been discussed. 
While X-ray crystal structures provide a glimpse of ordered water molecules interacting with the interior of GPCRs \citep{Yuan2013Angewandte,liu2012Science,Venkatakrishnan2013Nature,burg2015Science}, 
these structural waters are not completely static but have finite lifetimes.  
Furthermore, the roles of mobile water with a relatively faster relaxation kinetics remain unknown. 
Thus, it is timely to examine the dynamics of water in GPCR and ask the functional roles of water in conjunction with the local  dynamics of amino acid residues.  

Here, we explored water dynamics in A$_{\text{2A}}$AR by conducting $\mu$sec all-atom MD simulations.
In order to assess the role of water dynamics in GPCR function,
we strategically evaluated various quantities.
We first calculated the water capacity in the TM region, and mapped the probability density of water for each receptor state.
Explicit calculation of water flux through the TM region has revealed that the water flux in the agonist (UK432097)-bound active form is three times smaller than that in the apo or antagonist (ZM241385)-bound form.
Water streams are divided into multiple pathways, some of which form one-dimensional ``water wire" \citep{Raghavender2009JACS,Reddy2010PNAS,thirumalai2011ACR}. 
The receptor surfaces, mapped with the water relaxation time, indicate that the water dynamics is heterogeneous in space. 
Waters in the extra- and intra-cellular domains move fast ($\lesssim 1$ nsec). 
Especially in the agonist-bound form, waters trapped around the microswitches display ultra-slow relaxation ($\gtrsim 100$ nsec), coating the polar and charged surface of the TM channel.
Our study finally shows that water-mediated inter-residue interaction network, formed along microswitches in TM7, reinforces and extends the range of allosteric interface in the active state, aiding robust activation of GPCRs.

\section*{MATERIALS AND METHODS}

\noindent{\bf Preparation of the receptor structures complexed with ligands.}
The X-ray crystal structures of A$_{2A}$AR bound with an agonist or antagonist were modeled based on the structures from the Protein Data Bank (PDB) \citep{xu2011Science,jaakola2008Science}.
We used the X-ray crystal structures with PDB entries of 3QAK \citep{xu2011Science} for agonist-bound and 3EML \citep{jaakola2008Science} for antagonist-bound model, and used UK432097 and ZM241385 for agonist and antagonist ligands, respectively.
Since some of the loop regions are not resolved in the crystal structures, we performed homology modeling using the MODELLER program implemented in Discovery Studio v.3.1 to prepare the full-length A$_{2A}$AR models including all the loop regions.
The X-ray crystal structures with PDB entries of 2YDV \citep{lebon2011Nature} and 3PWH \citep{dore2011structure} were used to model the loop regions missing in 3QAK and 3EML, and the conserved disulfide bridges connecting the loops, i.e., C71-C159, C74-C146, C77-C166, and C259-C262, were retained.
These models were optimized by simulated annealing and selected based on the DOPE score.
The final structures were obtained by energy-minimization using the Conjugate Gradient method and the Generalized Born with simple SWitching (GBSW) implicit solvent model.

Although an antagonist-bound engineered crystal structure of A$_{\text{2A}}$AR complexed with apocytochrome b$_{562}$RIL replacing intracellular loop 3 (ICL3) (PDB entry: 4EIY \citep{liu2012Science}), whose structure was fully determined with no missing region including extracellular (EC) and intracellular (IC) loops, became available after we started this research,
the extent of structural overlap between the crystal structure and our modeled structure is as high as $C_{\alpha}$-RMSD $= 0.407$ \AA.
Furthermore, as the memory of initial configuration, especially the initial configuration of ``flexible" loops, would be erased after some time (at most a few tens of nsec), and an ``identical force field" is used for simulation, 
the conclusion of this study would not be altered even when a more precise and accurate structure is used as a starting conformation for the simulation run.
For the apo form, we used the protein structure, minimized after removing the antagonist ligand from the original structure (3EML) as a starting conformation of our MD simulations.
\\

\noindent{\bf Molecular dynamics simulations.}
To construct the explicit membrane system, the transmembrane region of A$_{2A}$AR was predicted based on the Orientations of Proteins in Membranes (OPM) database and the hydrophobicity of the protein.
Using SOLVATE 1.0
(http://www.mpibpc.mpg.de/home/grubmueller/solvate), we first solvated the receptor structure with TIP3P water molecules
and removed the water molecules outside the receptor in the TM region.
Then, the 1-Palmitoyl-2-oleoylphosphatidylcholine (POPC) lipid bilayer ($88\times 91$ \AA$^2$ wide) was placed around the TM region of A$_{2A}$AR.
After aligning the initially solvated receptor structure and the bilayer system,
we removed the lipid and water molecules overlapping within the distance of 0.5 \AA\ from the receptor structure.
The prepared system was solvated with the water box, covering all the previous output molecules,
and neutralized with K$^{+}$/Cl$^{-}$ ions to make 150 mM salt concentration.
Our system contains 68,799 (apo, 13,549 water + 173 lipid), 70,052 (antagonist-bound, 13,551 water + 182 lipid), and 69,449 (agonist-bound, 13,552 water + 177 lipid) atoms in $85\times 88\times 99$ \AA$^3$ periodic box.

MD simulations were performed using NAMD package v.2.8 \citep{NAMD} with CHARMM22/CMAP force field \citep{buck2006BJ}.
The topology and parameter files for the ligands were generated by SwissParam web server which provide topologies and parameters for ligand, based on Merck Molecular Force Field (MMFF), compatible with the CHARMM force field  \citep{Zoete2011JComputChem}.
The simulations were then conducted with the following steps:
(i) energy-minimization for 2,000 steps using conjugate gradient method in the order of membrane, water molecules, protein structure, and the whole system;
(ii) gradual heating from 0 K to 300 K using a 0.01 K interval at each step;
(iii) 50 nsec pre-equilibration with NVT ensemble before the production runs;
and
(iv) 1.2 $\mu$sec production runs with NPT ensemble ($T=300$ K, $P=1$ atm).
Pressure was maintained at $P = 1$ atm using the Nos{\'e}-Hoover Langevin piston method with an oscillation period of 200 fsec and a decay time of 50 fsec.
The temperature was controlled by applying Langevin forces to all heavy atoms with a damping constant of 1 psec$^{-1}$.
Non-bonded interactions were truncated with a 12.0 \AA\ cutoff and a 10.0 \AA\ switching distance, and the Particle Mesh Ewald summation with a grid spacing 0.91 \AA\ was used to handle electrostatic interactions.
We used a uniform integration time step of 1 fsec,
and during the whole production runs, no specific constraints were applied to the system except for restraining the hydrogen atoms.
SETTLE algorithm was used to keep water molecules rigid, and SHAKE to restrain the bond lengths of hydrogens other than water molecules.
The atomic coordinates of the simulated system were used every 50 psec for water flux analysis and 400 psec for other calculations.

\begin{figure}[ht]
\centering
\includegraphics[width=\columnwidth]{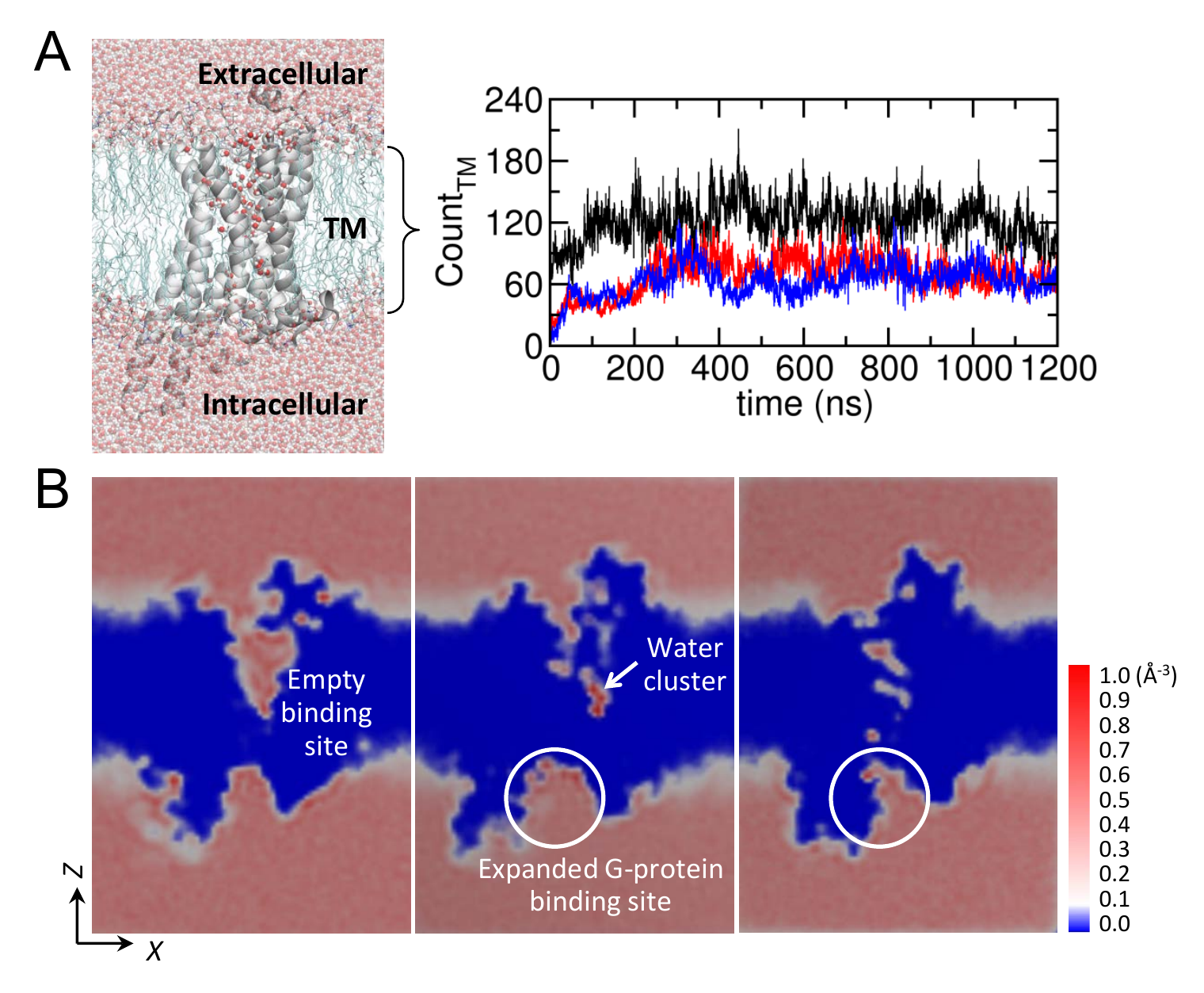}
\caption{
\small
{\bf Water molecules in the TM region.}
(A) The number of waters occupying the TM regions as a function of time in the apo (black), agonist-bound (red), and antagonist-bound (blue) forms.
On the left depicted is a snapshot of water permeating through the TM region.
(B) Probability density map of water $\rho({\bf r})$ at each 3D grid point ${\bf r}$ ($0\leq\rho({\bf r})\Delta {\bf r}\leq 1$ with the grid cell volume of $\Delta {\bf r}=\Delta x\Delta y\Delta z=$ 1 \AA\ $\times$ 1 \AA\ $\times$ 1 \AA) calculated for the apo, agonist-bound, and antagonist-bound states based on the final 200 nsec of simulation.
Two dimensional slices of $\rho({\bf r})$ were calculated using VolMap plugin in VMD.
\label{average_water_density}}
\end{figure}

Importantly, in the production runs it takes more than 200 -- 300 nsec to fully equilibrate the (receptor+lipid bilayer) system in NPT ensemble.
Water molecules fill the TM channel gradually (Figure \ref{average_water_density}). 
While the protein RMSD gets to its steady state at $t\lesssim 200$ nsec (Figure~S1A), the area per lipid or the bilayer thickness is still under relaxation in the first 200 -- 300 nsec (see Figure~S1B and S1C).
In the steady state, the area per lipid (APL) for POPC molecule is $\sim$ $61\pm 2$ \AA$^2$ and bilayer thickness is $\sim$ 40 \AA, which compare well with the experimental values at $T=27^{\circ}C$ (APL $\sim 63\pm 2$ \AA$^2$, thickness $\sim 39\pm 1$ \AA) as well as with other simulation studies \citep{kuvcerka2011BBA,tsai2013IJMS}.
After the first $t=0-300$ nsec simulation, we restrained the xy-plane area of the simulation box by using the option of `useConstantArea' in NAMD package. As a result, APL and thickness of lipid bilayer are stably maintained for $t\geq 300$ nsec. 
In our study, the production runs of the first 0 -- 400 nsec, which could usually be regarded a long simulation time, are still in the process of nonequilibrium relaxation.
Thus, we conducted various analyses in this study using the data from $t\geq 400$ nsec after the system had reached the steady state.
\\

\noindent{\bf Time correlation function to probe water unbinding kinetics.}
To quantitatively probe the dynamics of water near the receptor structure,
we calculated the auto-correlation function \citep{Luzar1996Nature,yoon2014JPCB}:
\begin{align}
C(t) &= \langle h(0)h(t) \rangle \nonumber\\
&= \frac {1}{N} \frac {1}{T-t} \int_{0}^{T-t} \sum_{i=1}^{N_{\alpha}} h_{i}(t'+t)h_{i}(t') dt'.
\end{align}
The function $h(t)$ describes whether a contact is formed between an atom in the receptor and a water molecule at time $t$.
$h(t)=1$ for a water molecule within 4 \AA\ from any heavy atom of a residue $\alpha$ at time $t$; otherwise, $h(t)=0$.
$\langle ... \rangle$
denotes a time average along the trajectory and an ensemble average over the number of water molecules ($N_{\alpha}$) associated with the residue $\alpha$.  
Since $h(0)=1$, $C(t)$ is equivalent to the survival probability of a bound water at time $t$.
Thus, the average lifetime of water is obtained using $\tau=\int^{\infty}_0dtC(t)$.
\\

\noindent{\bf Betweenness centrality.}
Simplifying a structure of complex system into a graph, represented with vertices and edges, can be used as powerful means to extract key structural features of the system \citep{freeman1979SocialNetworks,strogatz2001Nature,albert2000Nature,newman2005SocialNetworks}.
The graph theoretical analysis can be performed for protein structures \citep{greene2003JMB,amitai2004JMB,delSol2006MSB,bagler2007Bioinfomatics,vendruscolo2002PRE,dokholyan2002PNAS}, so as to judge the importance of each vertex that defines the graph.
For instance, a residue in contact with many other surrounding residues, which in the language of graph theory is called a vertex with high degree centrality, could be deemed important for characterizing a main feature of the given structure.
However, in order to address the issue of allostery, the betweenness centrality can be used for quantifying the extent to which a node ($\nu$) has control over the transmission of information between nodes in the network \citep{newman2005SocialNetworks}. 
The betweenness centrality is defined as
\begin{equation}
C_B(\nu)=\frac{2}{(N-1)(N-2)}\sum_{s=1}^{N-1}\sum_{t=s+1}^{N}\frac{\sigma_{st}(\nu)}{\sigma_{st}}
\label{CB}
\end{equation}
with $s,t\neq \nu$.
$\sigma_{st}$ is the number of paths with the shortest distance linking the nodes $s$ and $t$, and
$\sigma_{st}(\nu)$ is the number of minimal paths linking the nodes $s$ and $t$ via the node $\nu$. 

To build a network (graph), the residue-residue contact was defined (i) for water-free contact using the threshold distance of 4 \AA\ between any two heavy atoms of two residues, and (ii) for water-mediated contacts between two residues if a water oxygen is shared within 3.5 \AA\ by any two heavy atoms of two residues or if the two residues are within 4 \AA\ threshold distance as in (i).
To quantify the importance of the $\nu$-th residue in mediating signal transmission, 
we calculated two distinct betweenness centralities;  
(i) $C^o_B(\nu)$ for water-free and 
(ii) $C^{\text{w}}_B(\nu)$ for water-mediated inter-residue network of GPCR. 
$C^o_B(\nu)$ and $C^{\text{w}}_B(\nu)$ calculated for each snapshot of molecular structure were averaged over the ensemble of structures obtained along the MD trajectories.   
To calculate $C_B(\nu)$, we employed Brandes algorithm \citep{brandes2001JMS}, which substantially reduces the computational cost of Eq.\ref{CB}.

\section*{RESULTS}

\noindent{\bf Water capacity in the TM region.}
Our simulations show that it takes more than 200 -- 300 nsec for water molecules to fill the empty TM channel and to reach the steady state (Figure \ref{average_water_density}A,  see also Figure~S1 for the time evolutions of the receptor RMSD, area per POPC lipid, and membrane thickness).
In the steady state, on average 121$\pm$20, 69$\pm$18, and 63$\pm$16 water molecules are found in the interior (between the phosphate atoms of upper and lower leaflets) of the apo, agonist-bound active, and antagonist-bound inactive states, respectively (Figure \ref{average_water_density}A).
As depicted in the probability density map of water (Figure \ref{average_water_density}B),
the apo form, due to the water-filled ligand binding cleft, contains twice the volume of the water compared with the other ligand-bound states.
In the ligand-bound states, a large volume of water has to be discharged from the vestibule. 
The expanded G-protein binding cleft is also seen in the agonist-bound active state (Figure \ref{average_water_density}B, middle).
A water cluster, whose functional role will be further discussed in the Discussion, is identified in the midst of TM domain (see the white arrow in Figure \ref{average_water_density}B, middle).
\\

\begin{figure}[ht]
\centering
\includegraphics[width=0.8\columnwidth]{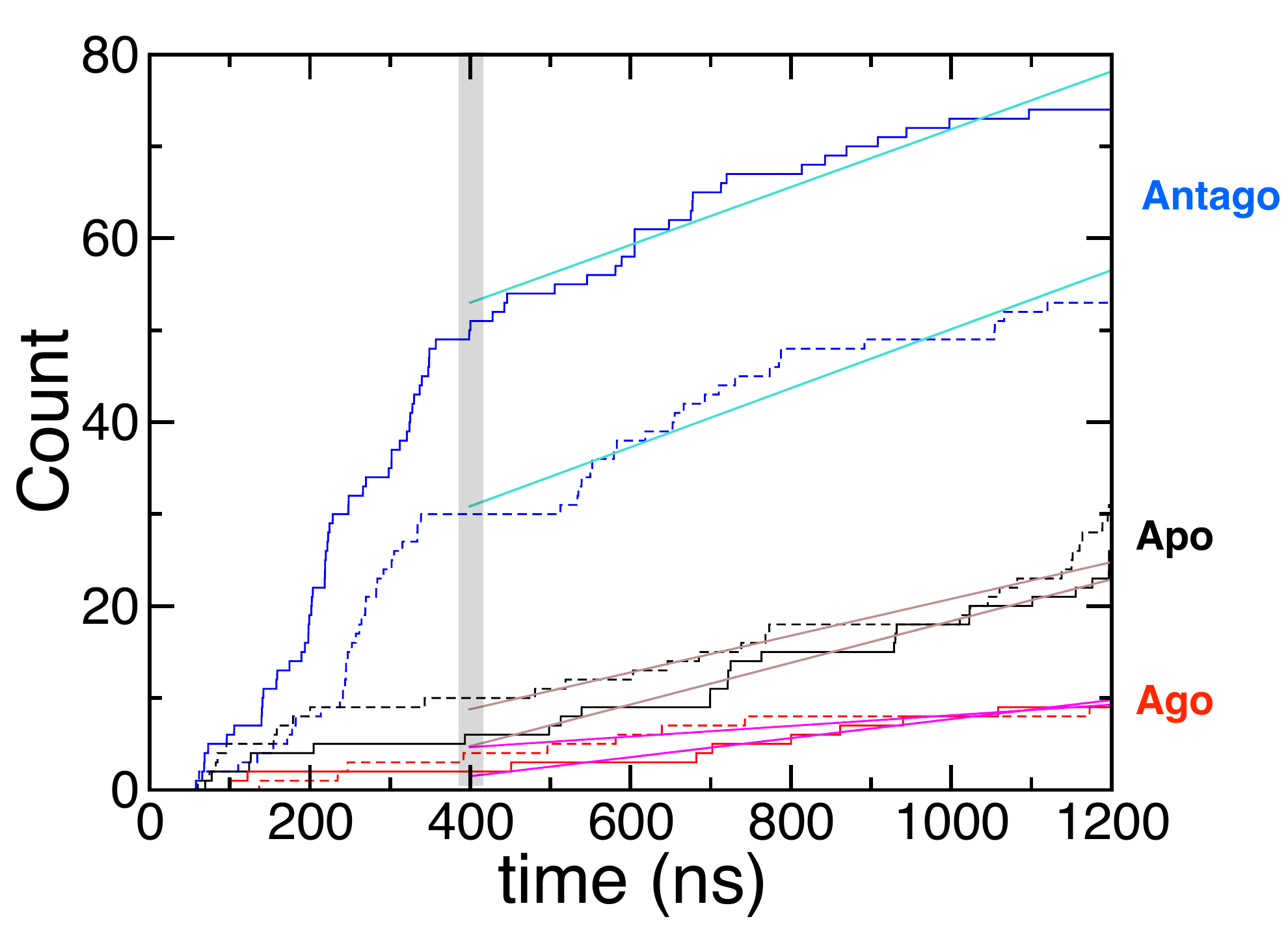}
\caption{
\small
{\bf Water flux through the transmembrane region.}
The number of water traversing TM domain as a function of time (EC$\rightarrow$IC: solid lines, IC$\rightarrow$EC: dashed lines).
Water fluxes ($j$) through the receptor, calculated using the slope from linear fits to the data over the interval $400<t<1200$ ns, are $j\approx 30$ (antagonist), $20$ (apo), and $10$ $\mu$sec$^{-1}$ (agonist).   
\label{flux}}
\end{figure}

\begin{figure*}[ht]
\centering
\includegraphics[width=0.9\textwidth]{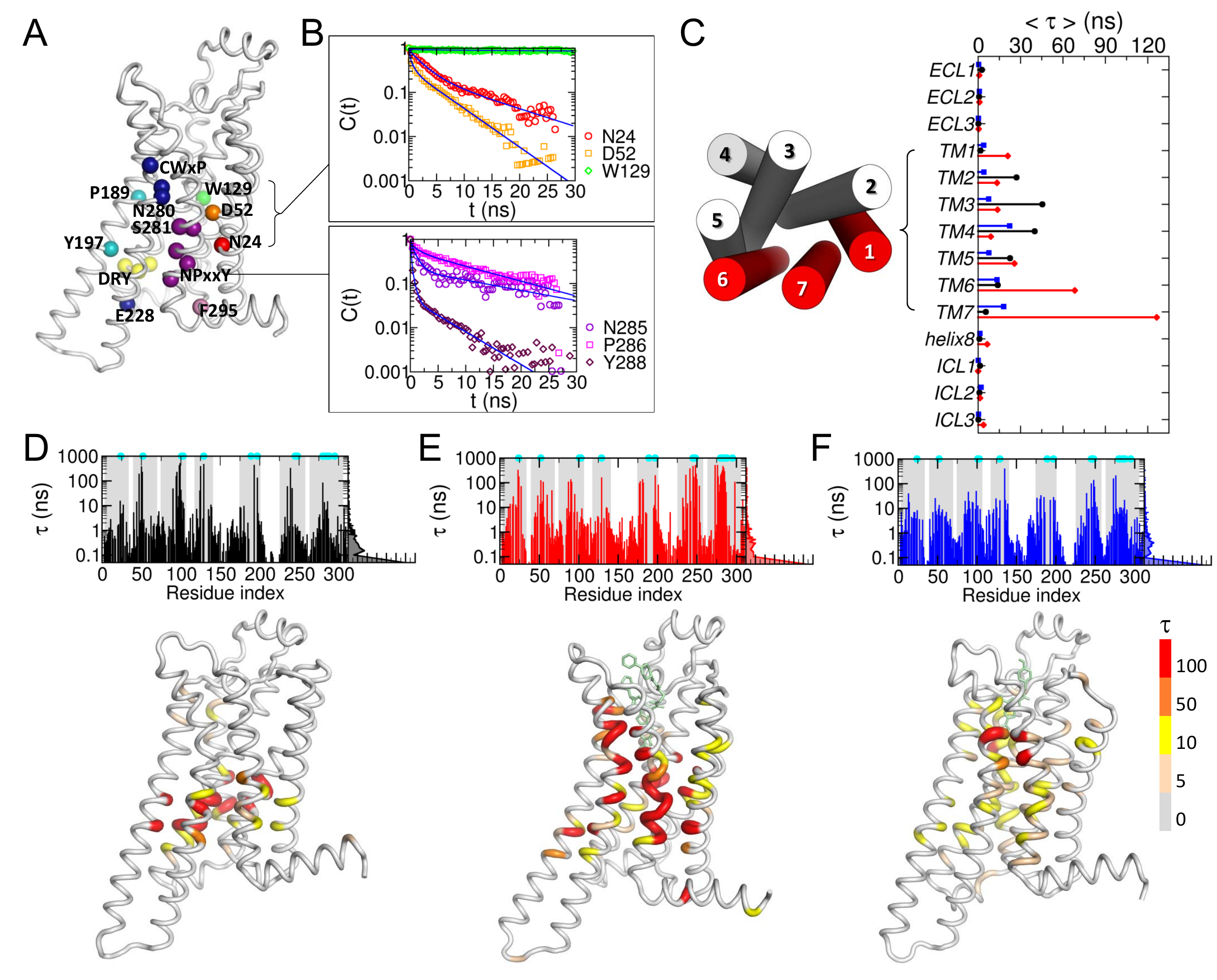}
\caption{
\small
{\bf Relaxation kinetics of water mapped on A$_{\text{2A}}$AR structure.}
(A) Location of the microswitches.
(B) Time correlation functions of water from six key residues are fitted to multi-exponential functions (see SI).
(C) Average relaxation time of water from various regions (TM, ECL, and ICL) in the apo (black), agonist-bound (red), and antagonist-bound (blue) forms.
The extracellular view of the TM helices is shown on the left. TM1, TM2, and TM7, which have much longer relaxation time in the agonist-bound form, are marked in red.
(D-F) Water relaxation time ($\tau$) from each residue in (D) the apo, (E) agonist-bound, and (F) antagonist-bound forms.
The TM regions are shaded in gray, and the positions of microswitches are marked with cyan dots.
The receptor structures are colored based on the $\tau$ values.
\label{corr_tau}}
\end{figure*}

\noindent{\bf Water flux through the TM pore.}
Although the average number of water in the TM channel remains constant in the steady state (Figure \ref{average_water_density}A, $t\gtrsim 400$ ns), this does not imply that water is static inside the channel.
The number of waters in and out of the channel are balanced in the steady state.
In order to compare the water flux between the different receptor states,
we traced the coordinate of every water molecule along the axis ($Z$-axis) perpendicular to the lipid bilayer plane 
and counted the number of water molecules that traverse through the channel from EC to IC or from IC to EC domain (Figure \ref{flux}).
In the early stage of simulations, non-steady state fluxes are observed in the apo and antagonist-bound forms.
Thus, we excluded the first 400 nsec from the analysis.
For $t>400$ ns, the water fluxes in the two directions satisfy $j_{\text{EC}\rightarrow\text{IC}}\approx j_{\text{IC}\rightarrow\text{EC}}$ in all the receptor states.
Overall, the water flux of the antagonist-bound state is three fold greater ($j_{\text{antago}}\sim 30$ $\mu$sec$^{-1}$) than that of the agonist-bound state ($j_{\text{ago}}\sim 10$ $\mu$sec$^{-1}$). The water flux of the apo state lies in between ($j_{\text{apo}}\sim 20$ $\mu$sec$^{-1}$).
In fact, the lowest water flux in the agonist-bound form has its molecular origin.
Below we will map each receptor state by probing the relaxation kinetics of water around each residue, which will help elucidate the role of water in receptor activation as well as the molecular origin of differential water fluxes.
\\

\noindent{\bf Relaxation kinetics of water.}
We mapped the water dynamics on the receptor surfaces by computing time correlation functions of water from each residue ($C(t)$, see Materials and Methods) (Figure \ref{corr_tau}A-B).
Water in the vicinity of the IC or EC loops, exposed to the bulk, is expected to have a shorter lifetime.
In the TM region, in contrast, dynamics of water is much slower (Figure \ref{corr_tau}C), displaying a broad spectrum of relaxation times depending on (i) the receptor state and (ii) their locations (Figure \ref{corr_tau}D-F).
The average lifetimes of water are 14.5 nsec, 26.3 nsec, and 7.1 nsec for the apo, agonist, and antagonist-bound states, respectively. 
The water molecules, hydrating the TM1, TM6, and TM7 helices in the agonist-bound form, reside longer than 100 nsec especially around the microswitches (marked with the cyan circles at the top of Figure \ref{corr_tau}D-F). 
This long residence time of water in the interior of TM domain ($\gtrsim\mathcal{O}(100)$ nsec) is noteworthy, given that a typical water lifetime on biomolecular surface probed by spin-label measurement is $\sim\mathcal{O}(100)$ psec \cite{franck2014JACS,franck2015JACS}. 

Similar conclusions as the above analysis calculating relaxation times were drawn by calculating the Fano factor involving water number fluctuation around each residue (see SI text and Figure~S2).

\begin{figure*}[ht]
\centering
\includegraphics[width=0.85\textwidth]{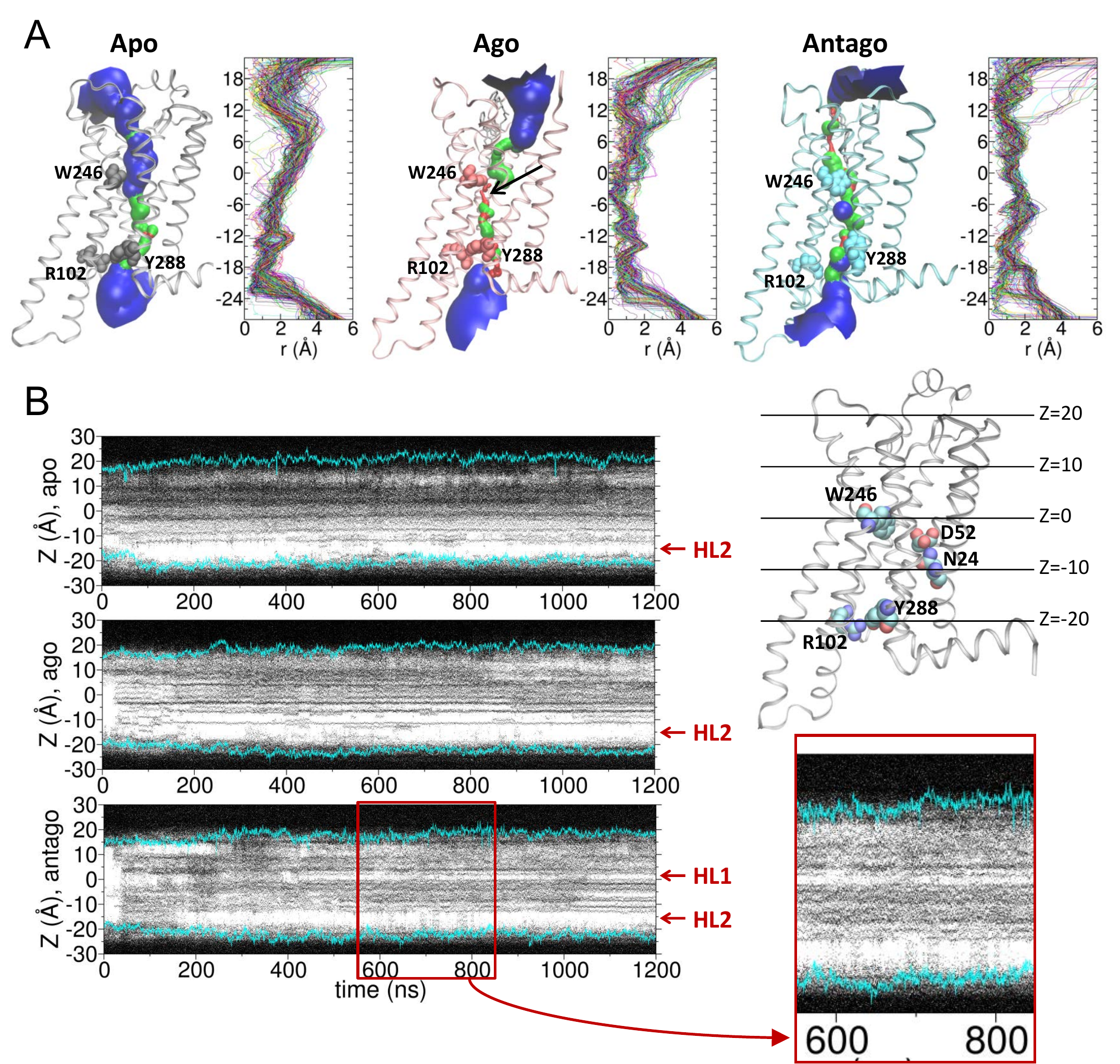}
\caption{
\small
{\bf Water occupation map of the TM channel.}
(A) The channel formed across the TM domain of each receptor state using the 250 snapshots from $1000-1200$ nsec for the apo form,
and from $600-800$ nsec for the agonist-bound and antagonist-bound forms, at which the highest water flux is observed.
The channel surfaces, calculated using the program HOLE \cite{Smart1996JMG}, are colored to visualize the channel radius (red ($r < 1.15$ \AA), green ($r = 1.15 \sim 2.30$ \AA), and blue ($r > 2.3$ \AA)).
Channel radii ($r$) along the $Z$-axis are shown on the right with different color for each frame.
(B) Water occupancy maps along the axis perpendicular to the bilayer, calculated for the simulations of the apo (top), agonist- (middle), and antagonist-bound forms (bottom).
The positions of two hydrophobic layers (HL1 and HL2) are marked with red arrows.
The position of the lipid headgroup is in cyan lines on the map.
A magnified view of water occupation map for $600\lesssim t\lesssim 800$ nsec is provided.
On the right, the apo structure is displayed with the key microswitches.
\label{water_flux_analysis}}
\end{figure*}

\section*{DISCUSSION}

\noindent{\bf Detailed look into water flux across the TM domain.}
In order to glean the molecular origin of receptor state-dependent water flux, we first visualize the geometry of water channel. 
The radius of water channel, calculated using the program HOLE \citep{Smart1996JMG} (see Figure \ref{water_flux_analysis}A), reveals that the geometrical bottleneck ($r\approx 0$) in the midst of TM domain ($Z\approx 0$) (Figure \ref{water_flux_analysis}A, see the black arrow) is formed around W246$^{6.48}$ and Y288$^{7.53}$ in the agonist-bound active form.
This is compatible with our finding that the water flux is substantially reduced in the agonist-bound form (Figure \ref{flux}).
W246$^{6.48}$, a key microswitch that senses an agonist and relays its signal to other microswitches \citep{Lee2015PLoSComp,Voth2013JACS}, is located deep inside the orthosteric binding vestibule, regulates the entry of water from the EC domain; whereas Y288$^{7.53}$ is located at the lower part of TM channel, regulating the entry of water from the IC domain.  

\begin{figure*}[t]
\centering
\includegraphics[width=0.9\textwidth]{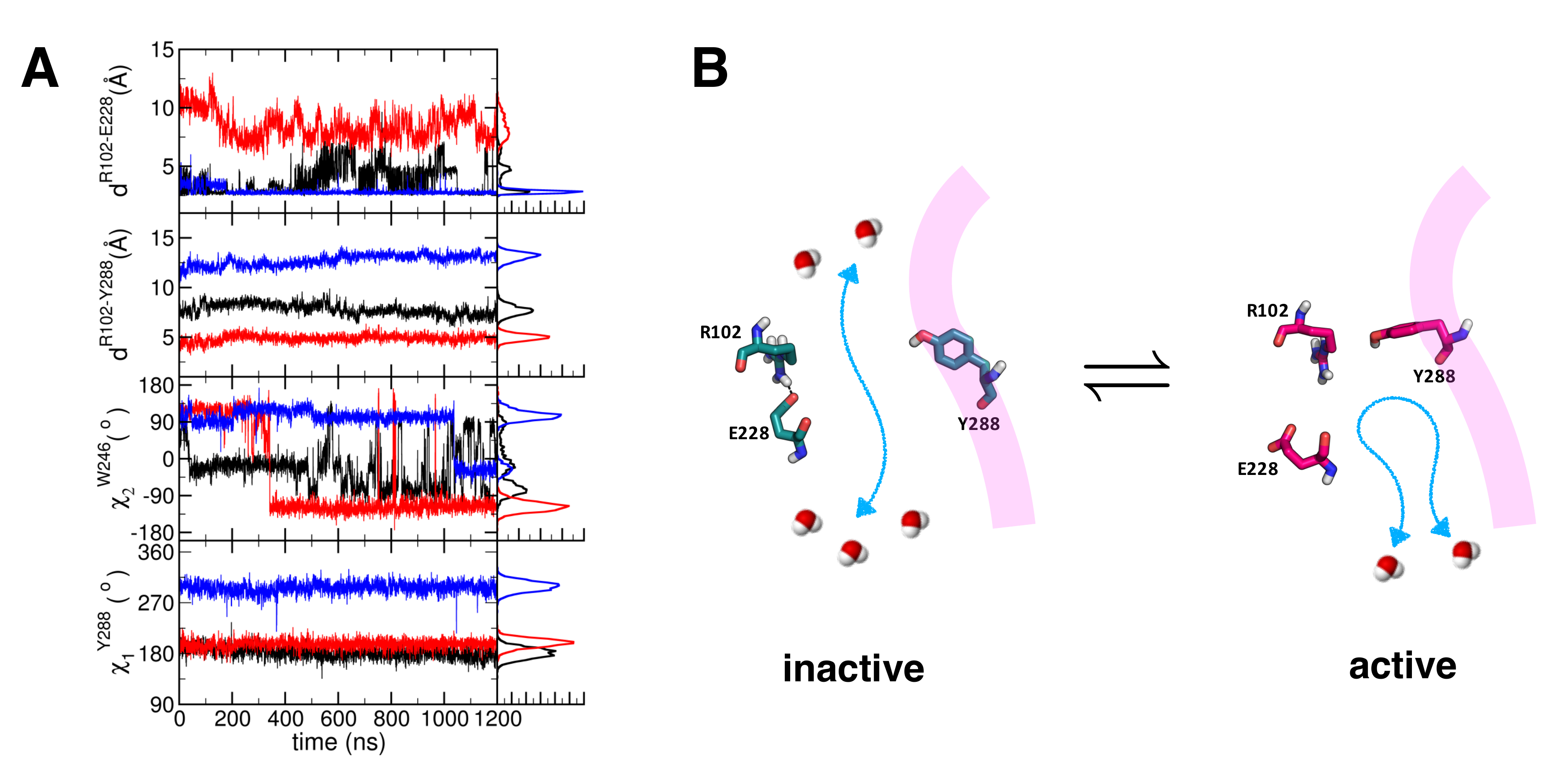}
\caption{
\small
{\bf Configuration of ionic-lock and Y288 associated with water flux gating.}
(A) Time traces probing the dynamics of three major structural motifs, i.e., ionic-lock (DRY motif), W246$^{6.48}$ of CWxP, and Y288$^{7.53}$ of NPxxY.
From the top shown are the distances between R102$^{3.50}$ and E228$^{6.30}$ (ionic-lock), and between R102$^{3.50}$ and Y288$^{7.53}$,
and the dihedral angles of W246$^{6.48}$ and Y288$^{7.53}$
for the apo (black), agonist-bound (red), and antagonist-bound (blue) forms.
The histograms on the right are drawn using data with $t>400$ ns.
(B) The configurations of R102$^{3.50}$, E228$^{6.30}$, and Y288$^{7.53}$ in the inactive and active states. The ionic lock (R102-E228) in the inactive form is disrupted in the active form, allowing R102 to form a contact with Y288. The passage of water from the IC domain is blocked by this change in the active form. The helix 7 is illustrated with the transparent band in purple.
\label{three_switches}}
\end{figure*}

Although the geometry of channel in each receptor state provides a glimpse of pipeline across the TM region,
the actual water flux through the channel is not fully explained by the radii of the channel alone.
For example, the apo state, overall, has greater radii along the channel and does not have a particularly more restrictive geometrical bottleneck than those in the antagonist-bound state (Figure \ref{water_flux_analysis}A); yet the flux is smaller than the one observed in the antagonist-bound state.
Depending on the extent of hydrophobicity or electrostatic nature of residues comprising each region of the channel surface,
stochastic wetting-dewetting transition \citep{aryal2014NatComm,Anishkin2004BJ,Reddy2010PNAS} can occur along the channel.
Furthermore, a stable water cluster in the channel, not exchanging water molecules with the surroundings, 
would impede the water dynamics through the channel.
Explicit calculations of water occupancy along the $Z$-axis \citep{Anishkin2004BJ} visualize how water molecules actually fill the TM channel at time $t$ (Figure \ref{water_flux_analysis}B).
As expected, in the apo state (Figure \ref{water_flux_analysis}B, top) the EC region ($Z>0$ \AA) is always filled with a high density of water since the empty ligand binding pocket is accessible from the bulk;
however, IC region ($-20\lesssim Z\lesssim -10$ \AA) remains ``dry", indicating that the entry of water through the IC region is blocked.
This dry zone corresponds to the second hydrophobic layer (HL2) around NPxxY motif above Y288$^{7.53}$, illustrated by Yuan \emph{et al.} \citep{Yuan2014NatCommun}.
In the agonist-bound form, another water-free layer is observed right above the HL2 ($-10 < Z < -5$ \AA) (Figure \ref{water_flux_analysis}B, middle).
This is the region below which the water cluster is formed.
On the other hand, the first hydrophobic layer (HL1) \citep{Yuan2014NatCommun}, corresponding to another dry zone between the orthosteric and the allosteric sites \citep{Yuan2014NatCommun}($Z\approx 0$ \AA) is observed in the inactive state.
Notably, despite the HL1, the water occupancy map of the antagonist-bound form (Figure \ref{water_flux_analysis}B, bottom) finds multiple instances ($t=600-800$ ns) that both HL1 and HL2 are filled with water bridging across the entire TM region. This accounts for the greater water flux in the inactive state.

\begin{figure*}[ht]
\centering
\includegraphics[width=1.7\columnwidth]{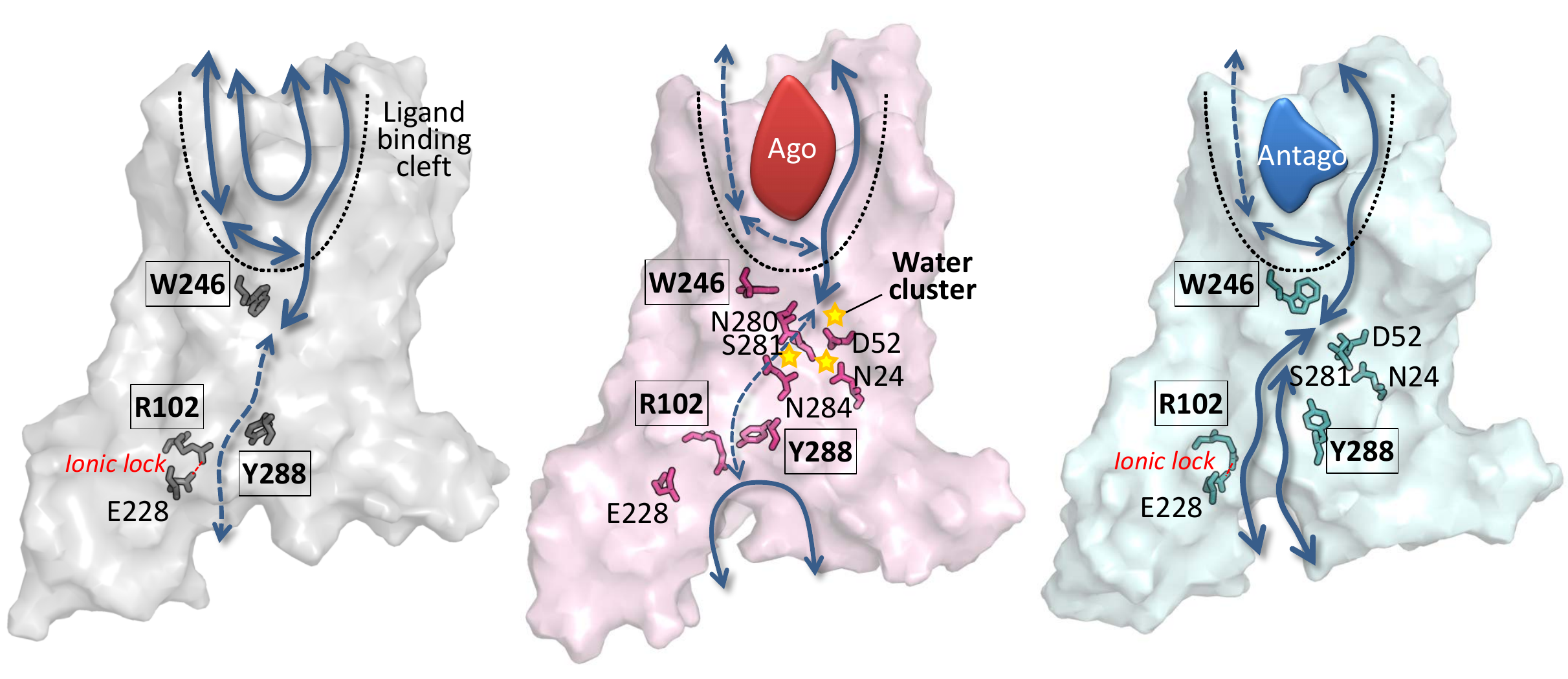}
\caption{
\small
{\bf Water flux map of A$_{\text{2A}}$AR.}
Schematics of water flux map (apo, agonist-bound, and antagonist-bound forms from the left to the right) drawn based on the water dynamics from SI movies, M1, M2, and M3.
The major and minor paths of water flux are depicted in solid and dashed lines, respectively.
The key microswitches are depicted in sticks, and the residues of the three major structural motifs, i.e., R102$^{3.50}$ of DRY, W246$^{6.48}$ of CWxP, and Y288$^{7.53}$ of NPxxY, are enclosed in the boxes.
In the agonist-bound form, the microswitches (N24$^{1.50}$, D52$^{2.50}$, N280$^{7.45}$, N284$^{7.49}$, S281$^{7.46}$) in TM1, 2, and 7 form water cluster and block the water flux.
The breakage of the ionic-lock brings R102$^{3.50}$ closer to Y288$^{7.53}$ in TM7, blocking the entry of water from  the IC domain.
\label{water_flux_map}}
\end{figure*}

The location of the hydrophobic layer and the receptor state-dependent water flux are affected by the alignment of polar residues along the channel.
The TM domain is mostly composed of non-polar hydrophobic residues, but there are polar/charged amino acids buried inside the TM domain as well.
The streams of water molecules are found along an ``Y" shaped array of these polar and charged residues that bridge through the TM domain (Figure~S3).
This array of polar/charged residues corresponds to the buried ionizable networks in GPCRs which have recently been underscored \citep{Isom2015PNAS}.
Our study reveals that these networks shape the passages of water molecules through the TM domain.
A misalignment of polar residues in the agonist-bound state is led to dehydration of the IC zone around $-20<Z<-10$ \AA, which gives rise to the HL2 (Figure~S3 left).
The rotameric state of Y288$^{7.53}$ sidechain in the antagonist-bound form enables a single file of water molecules constituting a water wire to flow across the HL2 (Figure~S3 right, SI Movie M3. See also Figure~S4).
The formation of the Y-shaped bridge made of polar residues including the NPxxY motif is the molecular origin underlying the ``hydrophobic gating" \citep{aryal2015JMB,powell2011NatureNanotech} that regulates the water flux through the IC region of A$_{\text{2A}}$AR.
\\

\noindent{\bf Microscopic origin of receptor state-dependent water flux.}
To further glean the microscopic underpinnings of the receptor state-dependent water flux (Figure \ref{flux}), configurations of microswitches at three key locations in the TM domain are probed (Figure \ref{three_switches}):
(i) The ionic lock (R102$^{3.50}$-E228$^{6.30}$) (Figure \ref{three_switches}A, top panel), the hallmark of the inactive state of GPCRs, is intact in the antagonist-bound form, maintaining the inter-residue distance  $d^{\text{R102-E228}}\approx 2.5$ \AA.
In the apo form, the ionic-lock repeatedly disrupts and rebinds, suggestive of the receptor's basal activity \citep{kobilka2007TPS}.
In the agonist-bound form, the ionic lock is completely disrupted;
(ii) The distance between R102$^{3.50}$ and Y288$^{7.53}$ depends on the receptor state (Figure \ref{three_switches}A, the second  panel from the top), and importantly Y288$^{7.53}$ gates the entry of water from the IC region. 
In the active form, R102$^{3.50}$ released from the influence of E228$^{6.30}$ can interact with and stabilize the side chain orientation of Y288$^{7.53}$, resulting in blocking the passage of water stream as well as misaligning the bridge made of NPxxY motif (see Supporting Movie M2, Figure \ref{three_switches}B, Figure~S3);
(iii) The side chain of W246$^{6.48}$ residue gates the entry of water from the EC domain.
In the active form, W246$^{6.48}$ blocks the water passage from the EC region, but it allows water to flow more freely in the inactive form.
In the apo form, the rotamer angle of W246$^{6.48}$ undergoes sharp transitions ($\chi^{\text{W246}}_2$ in Figure \ref{three_switches}A, black trace) multiple times during the simulation ($500\lesssim t\lesssim 1000$ ns), displaying correlations with the ionic-lock ($d^{\text{R102-E228}}$ in Figure \ref{three_switches}A, black line) and with the increased level of water flux (notice the sudden increase of the flux at $t\approx 500$ nsec from EC to IC (black trace in solid line in Figure \ref{flux})).
For example, when the ionic-lock was stabilized at $\approx 1000$ nsec in the apo form (Figure \ref{three_switches}A, top panel, black trace), the rotameric angle of W246$^{6.48}$ also displayed a sharp change from $-90^o$ to $+90^o$ (cyan arrow in the $\chi^{\text{W246}}_2$ plot of Figure \ref{three_switches}A).
This is the moment when $j_{\text{IC}\rightarrow \text{EC}}$ in the apo form has increased ($t>1000$ nsec in Figure \ref{flux}).

\begin{figure*}[ht]
\centering
\includegraphics*[width=1.6\columnwidth]{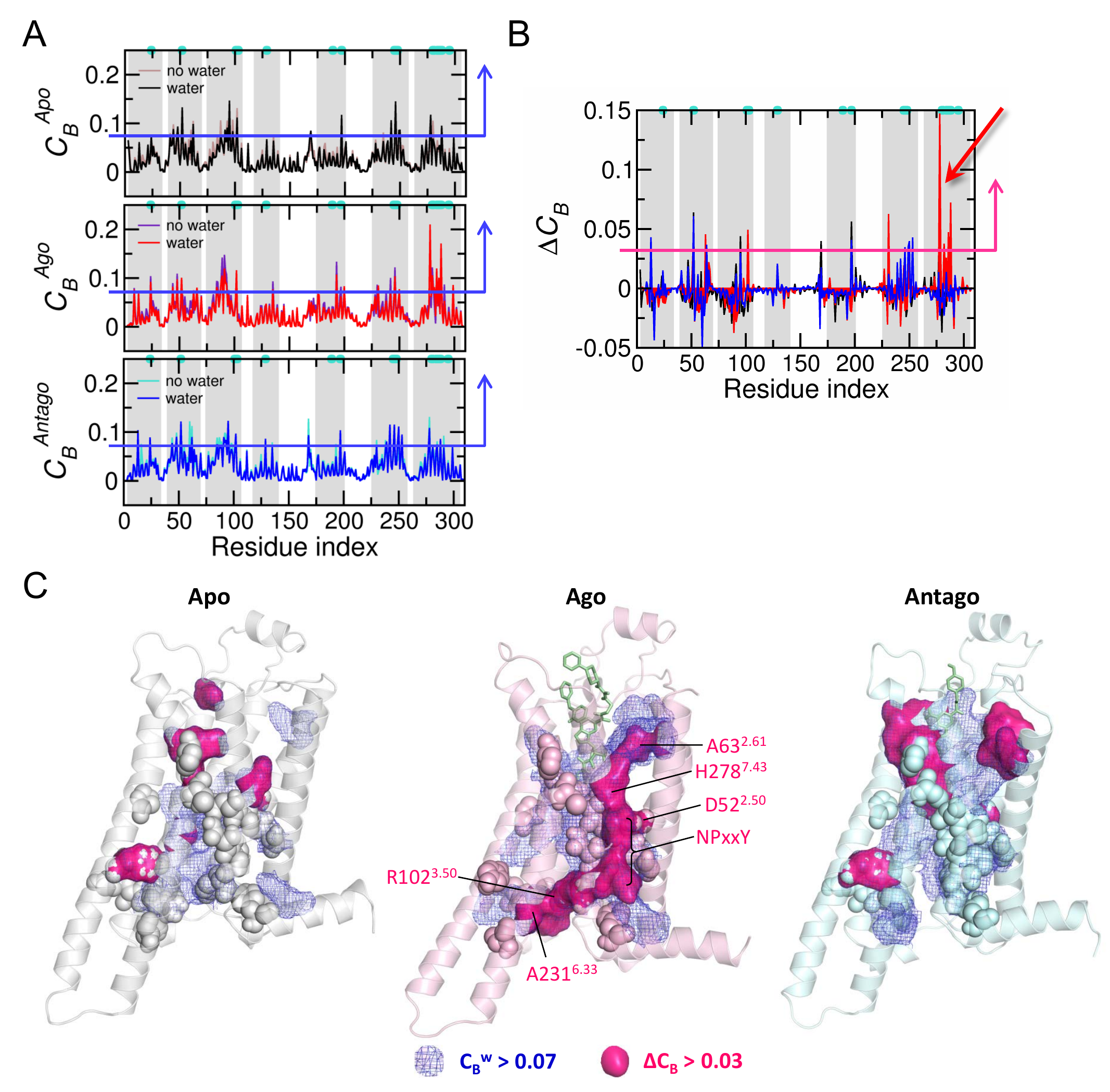}
\caption{{\bf Allosteric interface exteneded by water-mediated interactions.}
(A) Betweenness centralities calculated with ($C_B^{\text{w}}$) and without ($C_B^o$) water-mediated interaction for each receptor state.
Water-free contact between two residues is defined if any heavy atom in one residue is within 4 \AA\ from other residue; water-mediated contact between two residues is defined if a water oxygen is shared by two residues within 3.5 \AA.
(B) $\Delta C_B(\equiv C_B^{\text{w}}-C_B^o)$ for each state. Large values of $\Delta C_B$ are identified around TM7 helix in the agonist-bound active state (red arrow).
(C) Regions with $C_B^{\text{w}}>0.07$ (corresponding to the top 10 \% of $C_B$ values, blue mesh, and blue arrows in (A)) and $\Delta C_B>0.03$ (magenta surface in (C) and a magenta arrow in (B)) are demarcated on the structure of active state.
The microswitch residues are depicted in sphere representation.
\label{allosteric_interface}}
\end{figure*}

The movies (SI Movies M1, M2, and M3) from simulations of the water dynamics across TM region provide nice visualizations of receptor state-dependent water flux, which is recapitulated in the cartoons in Figure \ref{water_flux_map}.
In the apo form, the water molecules freely navigate the wide volume of the empty ligand binding cleft in the EC domain, but a further penetration across the TM region is regulated by the narrow channel gated by W246$^{6.48}$.
When the agonist or antagonist occupies the binding cleft, however, water flux is divided into the major (solid lines in Figure \ref{water_flux_map}) and minor streams (dashed lines in Figure \ref{water_flux_map});
the major stream is formed between TM1, 2, and 7, and the minor stream is formed between TM3, 5, and 6.
In the agonist-bound form, the W246$^{6.48}$ blocks the minor stream and the water flow along the major stream is tightly regulated by the several other microswitch residues (N24$^{1.50}$, D52$^{2.50}$, N280$^{7.45}$, S281$^{7.46}$, and N284$^{7.49}$), which creates the stable water cluster.
In case of the antagonist-bound form, no water cluster is observed; W246$^{6.48}$ gate is open and lets water molecules flow in and out of the TM channel. 
The water flux from the IC region is regulated by the side chain configuration of Y288$^{7.53}$. 
In the active state whose ionic-lock is disrupted, R102$^{3.50}$ interacts with Y288$^{7.53}$, which in turn blocks the passage of water flux in the IC domain.
\\

\noindent{\bf Allosteric interface reinforced by water-mediated interactions.}
In order to underscore the contribution of water-mediated contacts between TM residues to the receptor's allosteric signaling and function we conducted a graph theoretical analysis on the ensemble of GPCR structures.
As shown by our previous study \citep{Lee14Proteins} the microswitches of GPCRs in general are identified by graph theoretical analysis using betweenness centrality to be the key sites for intra-molecular orthosteric (allosteric) signal transmission (see Methods).
Allosteric interface can be visualized by highlighting those allosteric hotspots \citep{Lee14Proteins}.
While waters inside channel are generally dynamic, some water molecules, especially around microswitches in the active state, display slow relaxation kinetics and even can be trapped for an extended amount of time ($\tau>\mathcal{O}(10^2)$ ns) (see SI text, Figure~S5 and Figure~S6).
As long as water dynamics is sufficiently slow, stable water-mediated contacts can be made between two residues that are not in direct contact.
Defining a water-mediated contact when two residues share a water oxygen within 3.5 \AA\ from any heavy atom in each residue, we constructed a water-mediated residue interaction network for a given structure.
Using an ensemble of structure obtained from simulations, we calculated an average betweenness centrality ($C_B(\nu)$) at the $\nu$-th residue (see SI text, Figure \ref{allosteric_interface}A).  
At present, there are many ways to consider the protein allostery; some of them consider  thermodynamic aspect of protein allostery \cite{Hilser12ARB,Motlagh2014Nature}, and others focus more on identification of allosteric hotspot of a given structure \cite{Lockless99Science,halabi2009cell,Zheng05Structure}. 
The graph theoretical method \cite{Lee14Proteins,ribeiro2014JCTC,feher2014COSB,diPaola2015COSB} can also be employed to identify an allosteric hotspot of a given network structure, and in this method a residue with high $C_B$ value corresponds to a site important for allosteric signal transmission \citep{Lee14Proteins}.

In the agonist-bound active state, the $C_B(\nu)$ values calculated with ($C_B^{\text{w}}(\nu)$) and without ($C_B^{o}(\nu)$) water-mediated contacts (see Figure~S7) show clear differences ($\Delta C_B(\nu)$) along the microswitches in TM7 (Figure \ref{allosteric_interface}B, C), which is in accord with our findings that a number of slow water molecules stably coordinating with microswitches along the water channels are present in the active state.
Highlighted in Figure \ref{allosteric_interface}C with magenta surface is the allosteric interface in the active state reinforced by water-mediated interactions ($\Delta C_B(\nu)>0.03$). 
The interface mainly formed from along the TM7 helix, reaches R102$^{3.50}$ in the TM3 helix via Y288$^{\text{7.53}}$ and A231$^{\text{6.33}}$, spanning the whole TM  domain. 
We surmise that this wide-spread interface across the TM domain enables a robust long-range ``signal transmission" (compare the map of agonist-bound active state with those for apo and antagonist-bound inactive states in Figure \ref{allosteric_interface}C).
Notably, recent calculation of energy tranport in homodimeric hemoglobin also underscores the importance of interface water cluster, substantiating our proposal of ultra-slow water mediated allosteric signaling \cite{leitner2016JPCB}.

\section*{CONCLUSION}
Systematic analyses of GPCR structures reveal the network of inter-TM contacts mediated by microswitches \citep{Venkatakrishnan2013Nature}.
Network analysis also put forward that these microswitches act as the hub of the intra-receptor signaling network of A$_{\text{2A}}$AR \citep{Lee14Proteins}, and analyses of MD simulation trajectories confirmed that dynamics of microswitches occurs in concert \citep{Lee2015PLoSComp}.
Here, we extended our analysis to the dynamics of water molecules traversing through the TM channel and investigated their role in allosteric (orthosteric) signaling.

As is well appreciated, water, an essential component of living systems, provides driving force for self-assemblies of biomolecules and 
enhances their conformational fluctuations \citep{Yoon13JACS,yoon2014JPCB,Reat2000PNAS,Tsai00BJ,Fitter1996PNAS}.
Without water, biomolecules cannot function \citep{Frauenfelder2009PNAS}.
Guided by the array of polar residues, water permeates inside the mostly dry and hydrophobic pore of GPCRs.
In the antagonist-bound inactive state, this array of polar residues is connected continuously from the EC to IC domain.
In accord with Yuan \emph{et al.} \citep{Yuan2014NatCommun,Yuan2015Angewandte},
our simulations confirm the presence of the two hydrophobic layers, HL1 and HL2; however, these hydrophobic layers are not static, but highly dynamic. 
Our study also confirms the Yuan \emph{et al.}'s finding \citep{Yuan2015Angewandte} that the water flux along the TM channel is regulated by the rotameric states of several microswitches (Figure \ref{water_flux_map}).
Among them, two key microswitches, W246$^{6.48}$ and Y288$^{7.53}$ act together to form an ``AND-gate" controlling the water flow.


There are also other MD simulation studies on GPCRs (e.g. rhodopsin \cite{Leioatts2014Biochemistry}) which propose that a continuous stream of internal water is important for GPCR activation. 
By counting the number of water molecules inside TM domain, 
Leioatts \emph{et al.} \cite{Leioatts2014Biochemistry} showed that upon an elongation of retinal an influx of water increases inside the TM domain rhodopsin. 
They found that the increase of hydration level was significant in the complex-counterion simulation of rhodopsin, but not in the dark-state. 
Their simulation results on water hydration in the dark-state of rhodopsin differ from A$_{2A}$AR in the inactive state in that only 20 -- 30 water molecules are allowed inside the TM domain during the 1.6 $\mu$sec simulation time \cite{Leioatts2014Biochemistry}.   
However, the hydration level in the complex-counterion form is consistent with our study on the active form of A$_{2A}$AR in terms of the number of water molecules that fills the TM channel. 
In our study, both active and inactive states of A$_{\text{2A}}$AR could accommodate approximately an equivalent amount of water molecules, 60 -- 100, at steady state (Figure \ref{average_water_density}A), but the water flux in the inactive form was found greater than that in the inactive form by three times (Figure \ref{flux}). 
Here, it is crucial to distinguish ``the flux of water across TM domain" (Figure \ref{flux}) from ``the number of water inside TM domain" (Figure \ref{average_water_density}). 
As described in \emph{Results}, the water flux across TM domain was calculated by tracing the individual water molecule and counting the number of waters that enter and exit from one side of the membrane to the other. It is not merely the increase of TM water with time. 
While the number of water molecule inside the dark-state of rhodopsin is smaller than that in the inactive state of A$_{2A}$AR, the water flux in the dark-state of rhodopsin could also be large.  
Our study puts more emphasis on the dynamic aspect of water molecules across TM domain in the \emph{steady state}, which leads us to further ask the questions of which residues are hydrated by slow water and how those slow waters contribute to the water-contact mediated allosteric network.

Our explicit calculations of water fluxes indicate that the continuous water stream can be formed in all three states (i.e., $j\neq 0$ in Figure \ref{flux}), but with differing degrees. 
Thus, we want to argue that what is more relevant for GPCR activation is the water-mediated contacts among the key allosteric residues than the existence of continuous water stream. 
It is easier to make a water-mediated contact if a water molecule is slower. 
In the agonist-bound active state, water molecules inside the TM channel are almost stagnant, displaying minimal flux (Figure \ref{flux}); they can stably hydrate the microswitches mainly along the TM7 helix. 
We show that water-mediated residue network extends from extracellular domain to the intracellular part of TM6 helix via TM3 helix (Figure \ref{corr_tau} and Figure \ref{allosteric_interface}).

The waters around TM microswitches, some of which constitute the water cluster, stabilize the relative orientation and distance between TM helices by bridging them together (Figure \ref{allosteric_interface} and Figure~S7).
Our study highlights the interactions of internal water with microswitches, which contribute to extending and reinforcing the allosteric interface of GPCRs (Figure \ref{allosteric_interface}). 
Our study puts forward that these interactions are especially critical for the functional fidelity of the GPCR activity. 
\\

\section*{Acknowledgments}
This work was supported by the grant from the National Leading Research Laboratory (NLRL) program (2011-0028885) funded by the Ministry of Science, ICT \& Future Planning and the National Research Foundation of Korea (to S.C.), and by the RP-Grant 2015 funded by Ewha Womans University (to S.C. and Y.L.).
We thank KIAS and KISTI Supercomputing Center for providing computing resources.


\clearpage 
\section*{SUPPORTING INFORMATION}


{\bf Details of multiexponential fit of the auto-correlation function of water in Fig. 3.}
$C(t)$s of water around the residues given in Fig. 3B are fitted to tri-expoential function as follows:
$C^{N24}(t) = 0.656e^{-t/1.883{\text{ ns}}} + 0.047e^{-t/9.453{\text{ ns}}} + 0.296e^{-t/9.464{\text{ ns}}}$;
$C^{D52}(t) = 0.331e^{-t/0.128{\text{ ns}}} + 0.349e^{-t/0.790{\text{ ns}}} + 0.320e^{-t/5.005{\text{ ns}}}$;
$C^{W129}(t) = 0.079e^{-t/0.016{\text{ ns}}} + 0.021e^{-t/44.108{\text{ ns}}} + 0.900e^{-t/499.998{\text{ ns}}}$;
$C^{N285}(t) = 0.108e^{-t/0.319{\text{ ns}}} + 0.438e^{-t/2.365{\text{ ns}}} + 0.454e^{-t/12.546{\text{ ns}}}$;
$C^{P286}(t) = 0.001e^{-t/0.001{\text{ ns}}} + 0.085e^{-t/0.182{\text{ ns}}} + 0.915e^{-t/15.547{\text{ ns}}}$;
$C^{Y288}(t) = 0.852e^{-t/0.199{\text{ ns}}} + 0.104e^{-t/1.082{\text{ ns}}} + 0.044e^{-t/4.677{\text{ ns}}}$, 
and the average relaxation time of water is obtained by using
$\tau=\int^{\infty}_0C(t)dt$.
\\

\noindent{\bf Number fluctuation of water molecules on the receptor surfaces.}
Water undergoes sharp transition from the bulk to interface.
Heterogeneity of biomolecular surfaces and conformational dynamics give rise to a number of different classes of interfacial water \cite{li2007JACS,wood2007PNAS,yoon2014JPCB,fogarty2014JPCB}.
Tempo-spatial variation in the water number and its fluctuation on the receptor surfaces can provide glimpses of the water motion at the interfaces.
Some regions of the receptor are persistently occupied by water molecules with small number fluctation, and others with large fluctuation.
To quantify the water fluctuation, we counted the number of water molecules as a function of time, $n(t)$, around each residue, and calculated their average ($\langle n\rangle$) and variance ($\langle (\delta n)^2\rangle$) (Fig. \ref{fluctuation_fano}).
As expected, both $\langle n\rangle$ and $\langle(\delta n)^2\rangle$ are larger around the loop regions (ICLs and ECLs) than near the TM helices (see Fig. \ref{fluctuation_fano}).
Taking the ratio between the two numbers, $\langle n\rangle$ and $\langle(\delta n)^2\rangle$, i.e., $F=\langle \delta n^2\rangle/\langle n\rangle$ (Fano factor), one can appraise the influence of the receptor surface on water number fluctuation
(Fig. \ref{fluctuation_fano}).

In calculation, we counted the number of water molecules ($n$) within 4 \AA\ from any heavy atom of each amino acid residue.
Using our 1.2 $\mu$sec simulations, we calculated the average number of water ($\langle n\rangle$),
variance ($\langle(\delta n)^2\rangle=\langle n^2\rangle  - \langle n \rangle^2$),
and the ratio between the twos, i.e., Fano factor $F=\langle \delta n^2\rangle/\langle n\rangle$.
Residues with suppressed water fluctuations ($F\ll 1$) are commonly found in the ICLs and ECLs of all the three receptor states, and especially in the central zone of TM domain of apo and agonist-bound forms.
Notably, the region where the water cluster is detected (Fig. 1B, middle) is surrounded by the residues with $F\ll 1$ (Fig. \ref{fluctuation_fano}B, black arrows in the bottom panel).
Most of the residues with enhanced water fluctuations ($F >2$) are found at the interface between the receptor and lipid bilayer, especially in TM1, 2, 5, and helix 8.
The apo and agonist-bound forms have residues in the TM region with $F<0.25$, suggestive of strong attractive interactions from TM residues, whereas the TM region of the antagonist-bound form (Fig. \ref{fluctuation_fano}C) is devoid of such residues.
\\

\noindent{\bf Internal waters trapped inside the receptor.}
Among the water molecules that show slow dynamics, some molecules are trapped inside the receptor and tightly bound to specific sites.
Tracing the position of the entire water molecules along the trajectories,
we systematically identified the trapped water molecules.
We divided the full trajectory of 1.2 $\mu$s to six intervals (i.e., every 200 ns) and calculated the RMSD of each water oxygen relative to the position in the final snapshot of each interval.
Since the average RMSD of trapped water is smaller than that of untrapped free water ($\approx$ 66 \AA),
it is not difficult to identify the trapped water molecules.

The apo form has two such positions ($P_1^{\text{apo}}$ and $P_2^{\text{apo}}$ in Fig. \ref{trapped}A).
In $P_1^{\text{apo}}$, a water molecule is trapped via tight H-bondings with W129 and S47 during the entire simulation time (1.2 $\mu$s) (Fig. \ref{trapped_dist}A).
In case of $P_2^{\text{apo}}$,
maintaining the H-bonds with Y197, L95, and A99, three water molecules alternate to be trapped (see the three plateau regions in the graph plotting the distances between $P_2^{\text{apo}}$ and three water molecules in Fig. \ref{trapped}A).
That is, one water molecule in the site is displaced by another water (Supporting Movie M4).
Among the residues interacting with the trapped waters, W129$^{\text{4.50}}$ and Y197$^{\text{5.58}}$ residues are the microswitches.

In the agonist-bound form, trapped waters were observed in three positions (Fig. \ref{trapped}B):
The water molecules bound in $P^{\text{ago}}_1$ form the H-bonding interactions with N24$^{1.50}$, V282, and F286, connecting the TM1 and TM7 helices. In fact, $P_1^{\text{ago}}$ site corresponds to the position where the water cluster with high density is identified in the water density map (Fig. 1B, middle);
The water in $P^{\text{ago}}_2$ interacts with C245, L249, and A273, mediating the interaction between the TM6 and TM7;
Two water molecules are bound simultaneously to $P^{\text{ago}}_3$, maintaining the H-bonds with T279, N280$^{7.45}$, V283, and N284$^{7.49}$.
In case of $P^{\text{ago}}_2$ and $P^{\text{ago}}_3$,
several water molecules compete to bind at the sites (Figs. \ref{trapped_dist}B-C).
Along with these trapped waters, the water cluster is observed in the pocket formed by TM1, TM2, TM3, and TM7,
especially interacting with N24$^{1.50}$, D52$^{2.50}$, N280$^{7.45}$, and S281$^{7.46}$.
These water molecules constitutes an extended allosteric interface between TM helices, helping the receptor activation.

In the antagonist-bound form, trapped water molecules are observed in two sites (Fig. \ref{trapped}C).
One ($P^{\text{antago}}_1$) is near the DRY motif, which forms the ionic-lock (i.e., salt-bridge between R102$^{3.50}$ and E228 observed in the inactive state).
The water molecules in $P^{\text{antago}}_1$ do not form a direct interaction with DRY motif but make a tight H-bonding network with T117 and R120.
The residue R120 also makes the H-bonds with D101 and Y112.
The other site $P^{\text{antago}}_2$ is identified among the microswitch residues in TM1, TM2, and TM7.
The water molecule in this site makes the H-bonds with N24$^{1.50}$, S281$^{7.46}$, and Y288$^{7.53}$,
and in the antagonist-bound state, N24$^{1.50}$ interacts with D52$^{2.50}$ in TM2 via H-bonds.

In each receptor state, water molecules help extending interactions of the residues belonging to different TM helices, thus stabilizing the interhelical configurations of GPCRs.\\

\noindent{\bf Formation of water wires.}
A closer look at the movies of water dynamics through the TM domain (SI Movies M1, M2, and M3) reveals the presence of ``water wires" in a few locations.
Especially in the antagonist-bound inactive state, water molecules are aligned into one dimensional arrays and move in concert, stabilized by the successive H-bonds \citep{Raghavender2009JACS,Reddy2010PNAS}.
Macro-dipole moment,
$\vec{\mathcal{M}}=\sum_{i=1}^Nq_i\vec{r}_i$, where $i$ denotes the index of atoms (oxygen or hydrogen) comprising water molecules and $N$ is the total number of atoms, was calculated for the waters in the entire TM domain, and to demonstrate the presence of water wires quantitatively we calculated \emph{regional} macrodipole moment, $\vec{\mathcal{M}}_{\mathcal{R}}=\sum_{i\in\mathcal{R},i=1}^{N_{\mathcal{R}}}q_i\vec{r}_i$ where $N_{\mathcal{R}}$ is the number of atoms composing water molecules in the region $\mathcal{R}=A$, B, or C specified in Figure~S5.
Excluding the early stage of simulation ($t<400$ ns) when the water flux is not yet in the steady state,
we find that $|\vec{\mathcal{M}}|\approx 20$  Debye and occasionally reaches $|\vec{\mathcal{M}}|>55$ Debye (Figure~S5).
In the regions above and below W246$^{6.48}$, where water wires are clearly observed, the regional macrodipole moments are $|\vec{\mathcal{M}}_A|\approx 9.9\pm 2.6$, $|\vec{\mathcal{M}}_B|\approx9.8\pm 2.8$, and $|\vec{\mathcal{M}}_C|\approx5.7\pm 2.5$ Debye.
It is noteworthy that the orientation of the regional macro-dipole moment of the water wire in the region A is opposite to those in B and C.
\\

\section*{SUPPORTING MOVIES}

\noindent {\bf Supporting Movie M1.}
Water dynamics in the \emph{apo} form during the time interval $t=1100-1150$ ns.
All the water oxygens are shown with small spheres in different colors.
The key micro-switch residues are depicted using the stick representation marked with their residue numbers.
\\

\noindent{\bf Supporting Movie M2.}
Water dynamics in the \emph{agonist}-bound form during the time interval $t=700-750$ ns.
The details of the representation are identical with Movie M1.
\\

\noindent{\bf Supporting Movie M3.}
Water dynamics in the \emph{antagonist}-bound form during the time interval $t=700-750$ ns.
The details of the representation are identical with Movie M1.
\\

\noindent{\bf Supporting Movie M4.}
Dynamics of two water molecules in the apo form around the $P^{\text{apo}}_2$ for the time interval $t=821-835$ ns.
The receptor structure is represented as white ribbon, and two water molecules that compete around $P^{\text{apo}}_2$ are depicted using spheres in cyan and green.

\bigskip

\clearpage
\section*{SUPPORTING FIGURES}

\setcounter{figure}{0}

\makeatletter
\renewcommand{\thefigure}{S\@arabic\c@figure}
\makeatother

\begin{figure}[h]
\centering
 \includegraphics[width=2.3in]{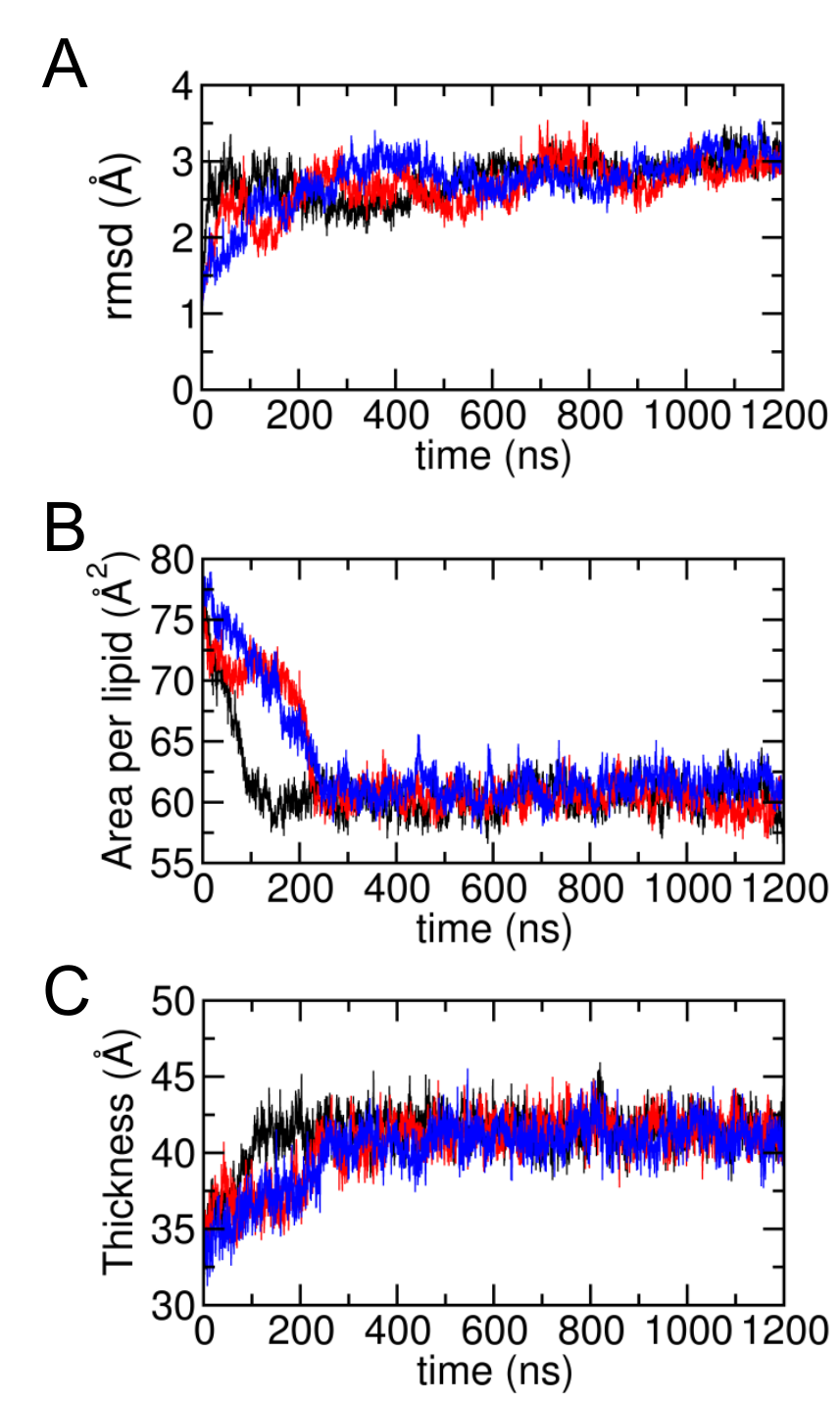}
  \caption{{\bf Protein RMSD and lipid characteristics during the simulations.}
(A) RMSD of the protein backbone structures for apo (black), agonist-bound (red), and antagonist-bound (blue) forms.
(B) Area per POPC lipid and (C) membrane thickness of bilayers (average distance between phosphorus atoms of POPC lipid in the upper and lower leaflets).
\label{system_equilibration}}
\end{figure}

\begin{figure*}[h]
\centering
 \includegraphics[width=5.6in]{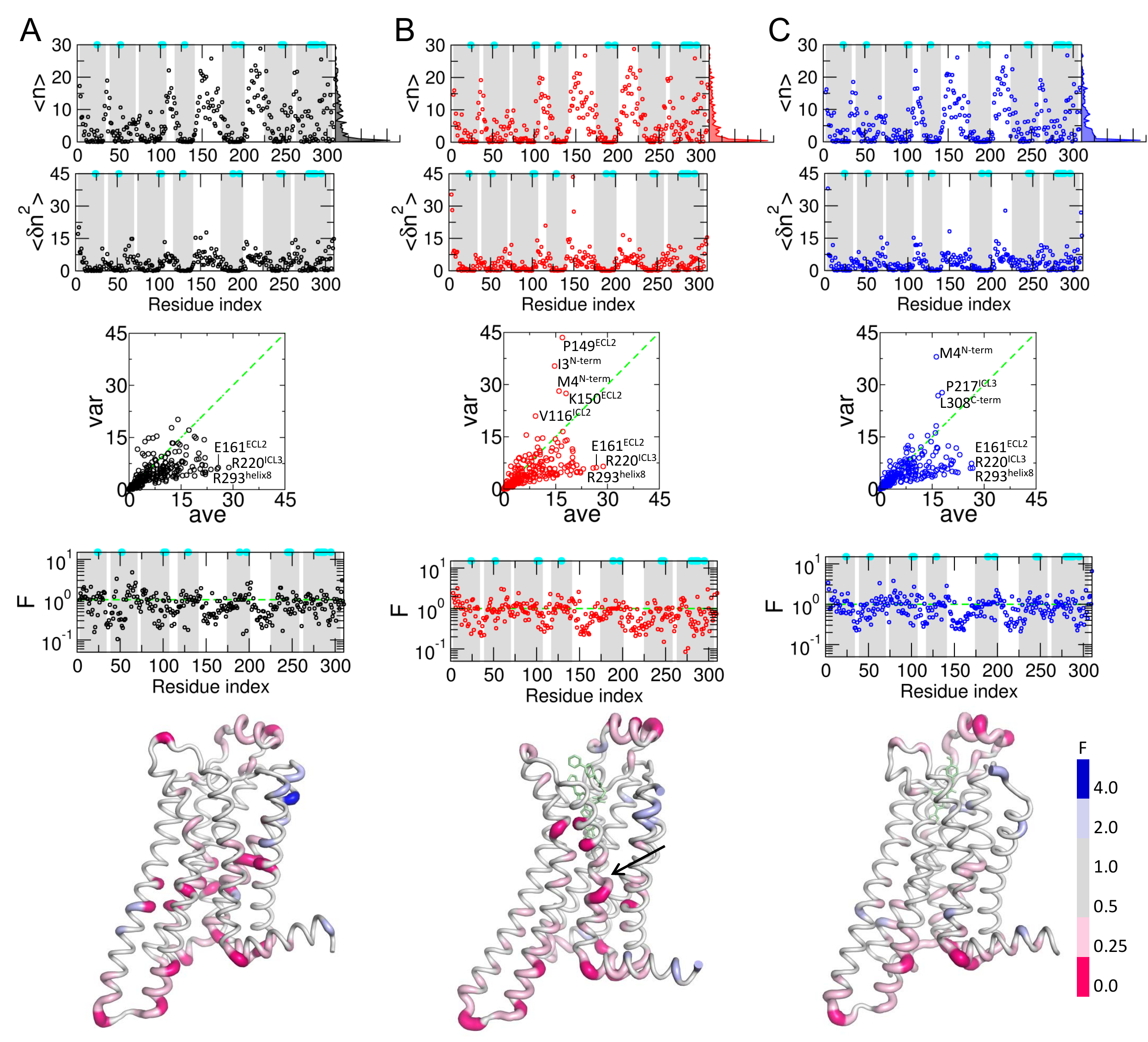}
  \caption{{\bf Number fluctuation of surface water molecules and their Fano factor.}
Average ($\langle n\rangle$) and variance ($\langle (\delta n)^2\rangle$) of water number are calculated around each residue using cut-off distance $R_c=4$ \AA\ from any heavy atom in the
(A) apo, (B) agonist-bound, and (C) antagonist-bound forms.
The ratio between the average and variance, i.e., Fano factor, are displayed at the bottom graph.
The TM regions are shaded in grey, and the positions of microswitch residues are marked with cyan dots.
The histogram of $\langle n\rangle$, $P(\langle n\rangle)$, over the residues are shown on the right hand side of the graph for $\langle n\rangle$.
In the scatter plots in the middle, the dashed line corresponds to $F=1$ ($\langle \delta n^2\rangle=\langle n\rangle$).
The residues with $F\gg 1$ are all found in the loop and terminal regions, and those with $F\ll 1$ and $25\lesssim \langle n\rangle < 30$ are commonly identified in all the ligand states to be E161, R220, R293 in the ECLs and ICLs.
The receptor structures are colored in accord with the \emph{F} value, and the bound ligands are shown in light-green sticks.
The region associated with the water cluster, identified in Fig. 1B, is marked with the arrow.
\label{fluctuation_fano}}
\end{figure*}

\begin{figure*}[h]
\centering
\includegraphics[width=1.5\columnwidth]{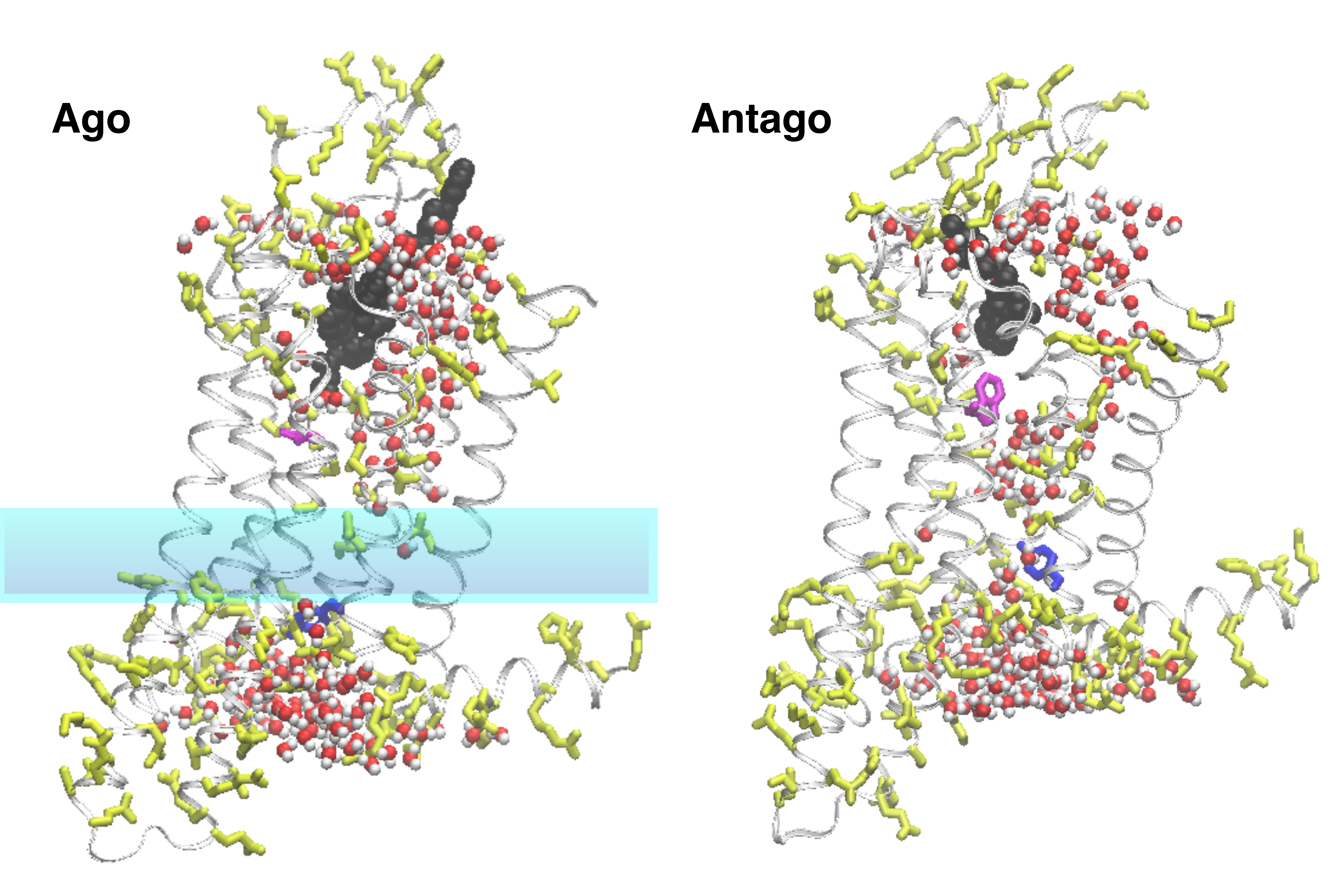}
\caption{
\small
{\bf Water pathways along the alignment of polar residues in the TM domain.}
TM domain is made mostly of nonpolar residues (white ribbon).
Polar residues, depicted together with their side-chain in yellow sticks, display a characteristic ``Y" shaped alignment, which internal water molecules in sphere representation hydrate.
The agonist and antagonist ligands are depicted in black spheres, and W246 and Y288 are shown in magenta and blue sticks, respectively.
In the agonist-bound form, the water free zone, corresponding to HL2, is highlighed in the middle, whereas in the antagonist-bound form, the water molecules are bridged through the TM channel along the polar residues.
\label{polar}}
\end{figure*}

\begin{figure*}[h]
\centering
\includegraphics*[width=5in]{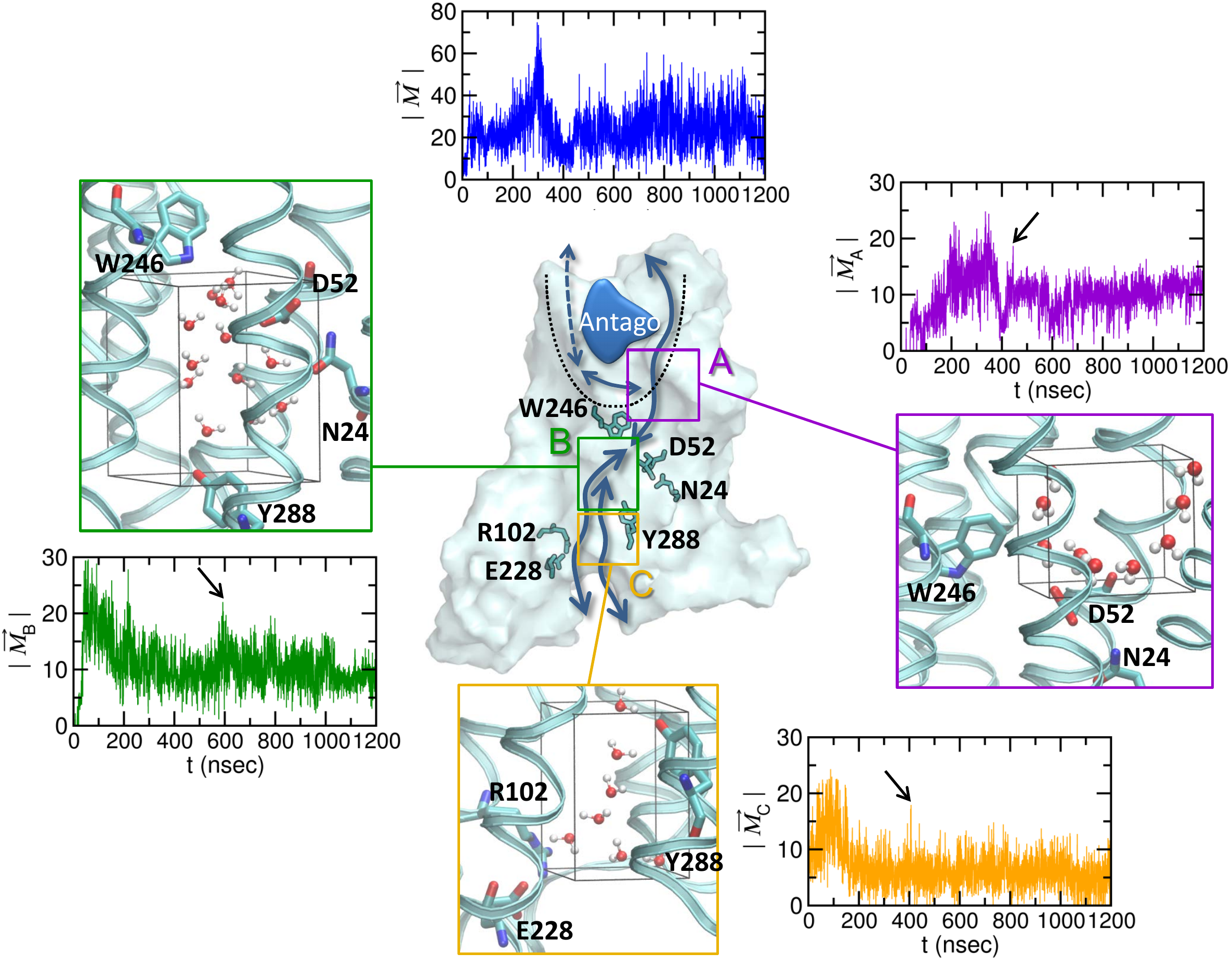}
\caption{
\small
{\bf Water wires.}
Macrodipole moment ($|\vec{\mathcal{M}}|$) for the entire population of water in the TM region (blue), and regional macrodipole moments ($|\vec{\mathcal{M}_{\mathcal{R}}}|$ with $\mathcal{R}=$ A, B, and C) calculated for the water in the regions A, B, and C.
The snapshots are taken for the water configuration in each region with the maximal regional macrodipole moments (marked with an arrow in each panel) for $t>400$ ns.
\label{water_wires}}
\end{figure*}

\begin{figure*}[h]
\centering
 \includegraphics[width=5.6in]{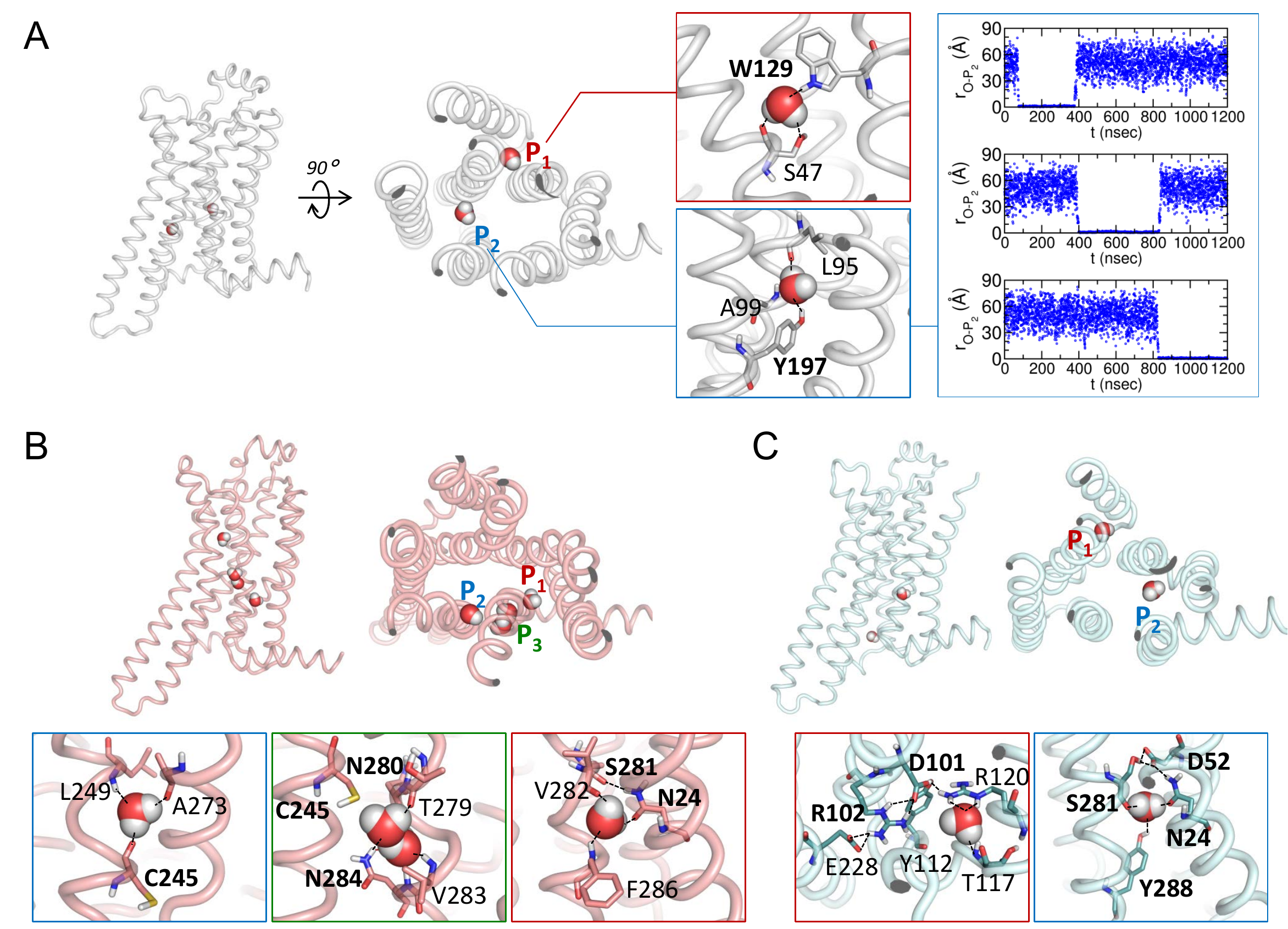}
  \caption{{\bf Trapped water molecules and their interactions with the receptor.}
(A) Trapped water molecules in the apo form.
The receptor secondary structure is displayed in tube, and the trapped water molecules are depicted in spheres.
The H-bonds are represented in black dashed lines, and the interacting residues are annotated.
The distances of the three water molecules relative to the site $P^{\text{apo}}_2$ are displayed on the right to show the water dynamics (see also Supporting movie M4).
(B) Agonist-bound form. (C) Antagonist-bound form. The distances of water molecules from the trapping sites are shown in Fig. \ref{trapped_dist}.
\label{trapped}}
\end{figure*}

\begin{figure*}[h]
\centering
 \includegraphics[width=4in]{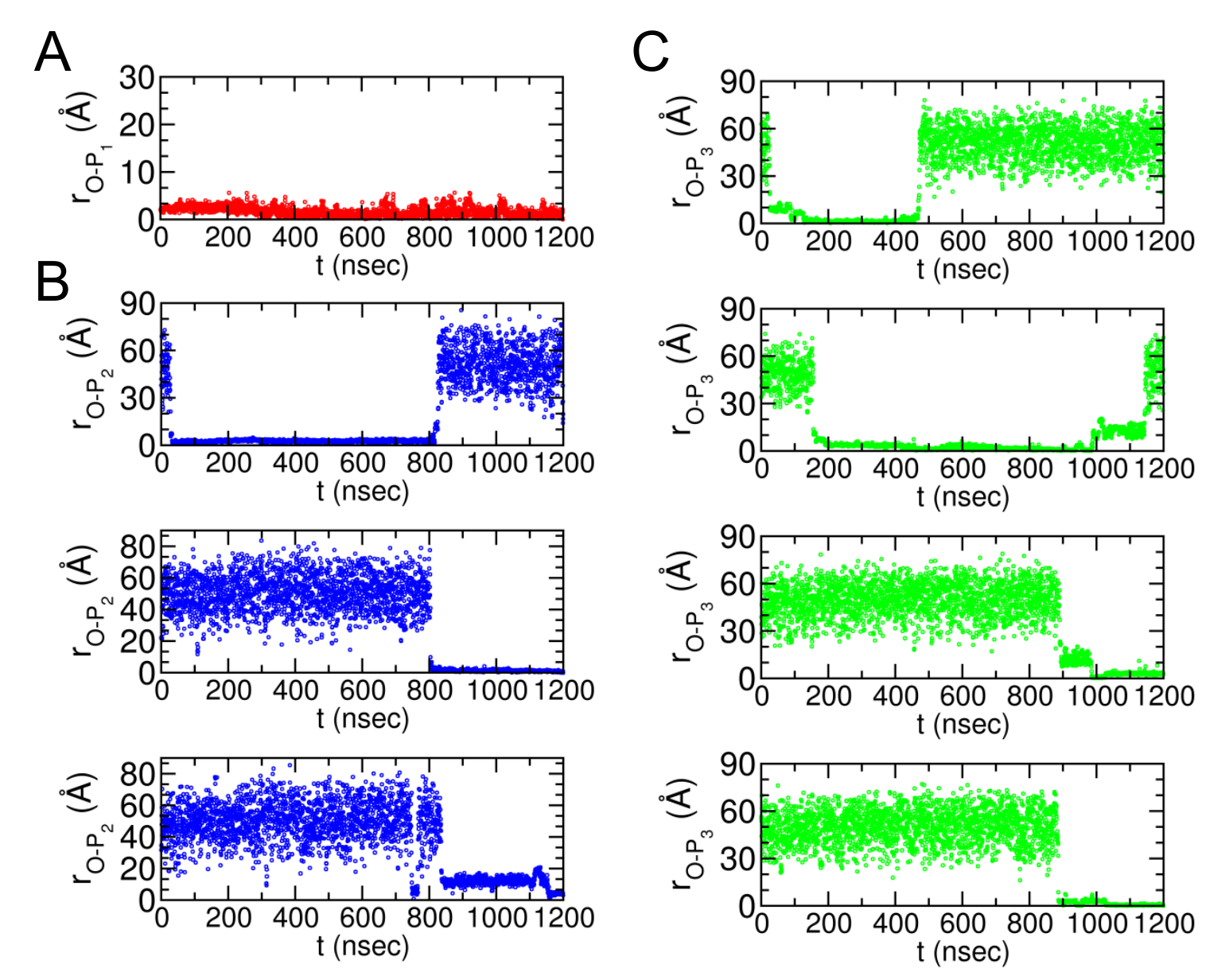}
   \caption{{\bf Dynamics of water trapping is visualized using the distance of a water molecule relative to the site where the water is trapped.}
(A) Dynamics of a water trapped in $P^{\text{apo}}_1$.
(B) Dynamics of three water molecules trapped in $P^{\text{ago}}_2$.
(C) Dynamics of four water molecules trapped in $P^{\text{ago}}_3$.
\label{trapped_dist}}
\end{figure*}

\begin{figure*}[h]
\centering
 \includegraphics[width=5.6in]{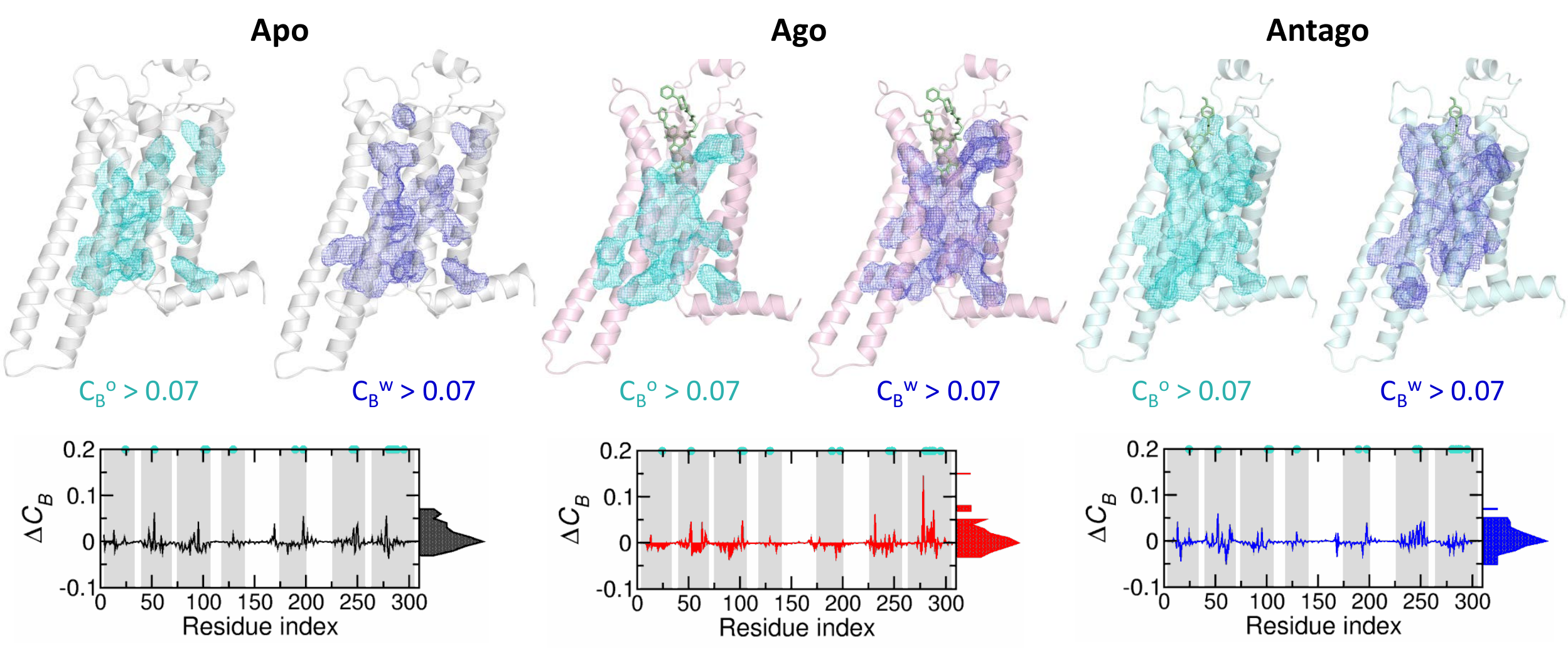}
  \caption{{\bf Analysis of allosteric interface using betweenness centrality.}
Betweenness centrality ($C_B$) were calculated for two residue interaction networks.
  (i) $C_B^o$, based on the network that takes into account the residue-residue interaction in direct contact (between any heavy atom in the residues within $R_c\leq 4$ \AA), and (ii) $C_B^{\text{w}}$, based on the network that takes into account the water-mediated residue-residue contact in which two residues either shares the same water molecule and satisfies residue-water oxygen distance cut-off value of 3.5 \AA\ or are in direct contact within the distance of 4 \AA.
  Allosteric interface using $C_B^o>0.07$ and $C_B^{\text{w}}>0.07$ are shown on the top for each receptor state.
\label{betweenness}}
\end{figure*}


\begin{thebibliography}{77}
\providecommand{\url}[1]{\texttt{#1}}
\providecommand{\urlprefix}{ }

\bibitem[Rosenbaum et~al.(2009)Rosenbaum, Rasmussen, and
  Kobilka]{Rosenbaum2009Nature}
Rosenbaum, D., S.~Rasmussen, and B.~Kobilka, 2009.
\newblock {The structure and function of G-protein-coupled receptors}.
\newblock \emph{Nature} 459:356--363.

\bibitem[Rasmussen et~al.(2011)Rasmussen, DeVree, Zou, Kruse, Chung, Kobilka,
  Thian, Chae, Pardon, Calinski, Mathiesen, Shah, Lyons, Caffrey, Gellman,
  Steyaert, Skinoitis, Weis, Sunahara, and Kobilka]{rasmussen2011Nature}
Rasmussen, S., B.~DeVree, Y.~Zou, A.~Kruse, K.~Chung, T.~Kobilka, F.~Thian,
  P.~Chae, E.~Pardon, D.~Calinski, J.~Mathiesen, S.~Shah, J.~Lyons, M.~Caffrey,
  S.~Gellman, J.~Steyaert, G.~Skinoitis, W.~Weis, R.~Sunahara, and B.~Kobilka,
  2011.
\newblock {Crystal structure of the $\beta_2$ adrenergic receptor-Gs protein
  complex}.
\newblock \emph{Nature} 477:549--555.

\bibitem[Lebon et~al.(2011)Lebon, Warne, Edwards, Bennett, Langmead, Leslie,
  and Tate]{lebon2011Nature}
Lebon, G., T.~Warne, P.~Edwards, K.~Bennett, C.~Langmead, A.~Leslie, and
  C.~Tate, 2011.
\newblock {Agonist-bound adenosine A2A receptor structures reveal common
  features of GPCR activation}.
\newblock \emph{Nature} 474:521--525.

\bibitem[Nygaard et~al.(2009)Nygaard, Frimurer, Holst, Rosenkilde, and
  Schwartz]{nygaard2009TPS}
Nygaard, R., T.~Frimurer, B.~Holst, M.~Rosenkilde, and T.~Schwartz, 2009.
\newblock Ligand binding and micro-switches in {7TM} receptor structures.
\newblock \emph{Trends Pharmacol. Sci.} 30:249--259.

\bibitem[Katritch et~al.(2013)Katritch, Cherezov, and
  Stevens]{katritch2013ARPT}
Katritch, V., V.~Cherezov, and R.~C. Stevens, 2013.
\newblock {Structure-function of the G protein-coupled receptor superfamily}.
\newblock \emph{Annu. Rev. Pharmacol. Toxicol.} 53:531--556.

\bibitem[Lee et~al.(2015)Lee, Choi, and Hyeon]{Lee2015PLoSComp}
Lee, Y., S.~Choi, and C.~Hyeon, 2015.
\newblock {Communication over the network of binary switches regulates the
  activation of A2A adenosine receptor}.
\newblock \emph{PLoS Comp. Biol.} 11:e1004044.

\bibitem[Lee et~al.(2014)Lee, Choi, and Hyeon]{Lee14Proteins}
Lee, Y., S.~Choi, and C.~Hyeon, 2014.
\newblock {Mapping the intramolecular signal transduction of G-protein coupled
  receptors}.
\newblock \emph{Proteins: Struct. Func. Bioinfo.} 82:727--743.

\bibitem[Tarek and Tobias(2000)]{Tarek00BJ}
Tarek, M., and D.~J. Tobias, 2000.
\newblock {The Dynamics of Protein Hydration Water: A Quantitative Comparison
  of Molecular Dynamics Simulations and Neutron-scattering Experiments}.
\newblock \emph{Biophys. J.} 79:3244--3257.

\bibitem[Ball(2008)]{Ball2008ChemRev}
Ball, P., 2008.
\newblock Water as an active constituent in cell biology.
\newblock \emph{Chem. Rev.} 108:74--108.

\bibitem[Frauenfelder et~al.(2009)Frauenfelder, Chen, Berendzen, Fenimore,
  Jansson, McMahon, Stroe, Swenson, and Young]{Frauenfelder2009PNAS}
Frauenfelder, H., G.~Chen, J.~Berendzen, P.~W. Fenimore, H.~Jansson, B.~H.
  McMahon, I.~R. Stroe, J.~Swenson, and R.~D. Young, 2009.
\newblock A unified model of protein dynamics.
\newblock \emph{Proc. Natl. Acad. Sci. U. S. A.} 106:5129--5134.

\bibitem[Tsai et~al.(2000)Tsai, A.Neumann, and Bell]{Tsai00BJ}
Tsai, A.~M., D.~A.Neumann, and L.~N. Bell, 2000.
\newblock {Molecular Dynamics of Solid-State Lysozyme as Affected by Glycerol
  and Water: A Neutron Scattering Study}.
\newblock \emph{Biophys. J.} 79:2728--2732.

\bibitem[Pal et~al.(2002)Pal, Peon, and Zewail]{Pal2002PNAS}
Pal, S.~K., J.~Peon, and A.~H. Zewail, 2002.
\newblock Biological water at the protein surface: dynamical solvation probed
  directly with femtosecond resolution.
\newblock \emph{Proc. Natl. Acad. Sci. U. S. A.} 99:1763--1768.

\bibitem[Jungwirth(2015)]{Jungwirth2015JPCL}
Jungwirth, P., 2015.
\newblock Biological Water or Rather Water in Biology?
\newblock \emph{J. Phys. Chem. Lett.} 6:2449--2450.

\bibitem[Mallikarjunaiah et~al.(2011)Mallikarjunaiah, Leftin, Kinnun, Justice,
  Rogozea, Petrache, and Brown]{Mallikarjunaiah2011BJ}
Mallikarjunaiah, K., A.~Leftin, J.~J. Kinnun, M.~J. Justice, A.~L. Rogozea,
  H.~I. Petrache, and M.~F. Brown, 2011.
\newblock Solid-state 2 H NMR shows equivalence of dehydration and osmotic
  pressures in lipid membrane deformation.
\newblock \emph{Biophys. J.} 100:98--107.

\bibitem[Yuan et~al.(2014)Yuan, Filipek, Palczewski, and
  Vogel]{Yuan2014NatCommun}
Yuan, S., S.~Filipek, K.~Palczewski, and H.~Vogel, 2014.
\newblock {Activation of G-protein-coupled receptors correlates with the
  formation of a continuous internal water pathway}.
\newblock \emph{Nat. Commun.} 5:4733.

\bibitem[Yuan et~al.(2013)Yuan, Vogel, and Filipek]{Yuan2013Angewandte}
Yuan, S., H.~Vogel, and S.~Filipek, 2013.
\newblock {The Role of Water and Sodium Ions in the Activation of the
  $\mu$-Opioid Receptor}.
\newblock \emph{Angew. Chemie International Ed.} 52:10112--10115.

\bibitem[Sun et~al.(2014)Sun, {\AA}gren, and Tu]{Sun2014JPCB}
Sun, X., H.~{\AA}gren, and Y.~Tu, 2014.
\newblock Functional Water Molecules in Rhodopsin Activation.
\newblock \emph{J. Phys. Chem. B} 118:10863--10873.

\bibitem[Burg et~al.(2015)Burg, Ingram, Venkatakrishnan, Jude, Dukkipati,
  Feinberg, Angelini, Waghray, Dror, Ploegh, and Garcia]{burg2015Science}
Burg, J.~S., J.~R. Ingram, A.~Venkatakrishnan, K.~M. Jude, A.~Dukkipati, E.~N.
  Feinberg, A.~Angelini, D.~Waghray, R.~O. Dror, H.~L. Ploegh, and K.~C.
  Garcia, 2015.
\newblock {Structural basis for chemokine recognition and activation of a viral
  G protein-coupled receptor}.
\newblock \emph{Science} 347:1113--1117.

\bibitem[Leioatts et~al.(2014)Leioatts, Mertz, Mart{\'\i}nez-Mayorga, Romo,
  Pitman, Feller, Grossfield, and Brown]{Leioatts2014Biochemistry}
Leioatts, N., B.~Mertz, K.~Mart{\'\i}nez-Mayorga, T.~D. Romo, M.~C. Pitman,
  S.~E. Feller, A.~Grossfield, and M.~F. Brown, 2014.
\newblock Retinal ligand mobility explains internal hydration and reconciles
  active rhodopsin structures.
\newblock \emph{Biochemistry} 53:376--385.

\bibitem[Lee and Lyman(2012)]{Lee2012BJ}
Lee, J.~Y., and E.~Lyman, 2012.
\newblock {Agonist Dynamics and Conformational Selection during Microsecond
  Simulations of the A2A Adenosine Receptor}.
\newblock \emph{Biophys. J.} 102:2114--2120.

\bibitem[Yuan et~al.(2015)Yuan, Hu, Filipek, and Vogel]{Yuan2015Angewandte}
Yuan, S., Z.~Hu, S.~Filipek, and H.~Vogel, 2015.
\newblock {W246$^{6. 48}$ Opens a Gate for a Continuous Intrinsic Water Pathway
  during Activation of the Adenosine A$_{2A}$ Receptor}.
\newblock \emph{Angewandte Chemie} 127:566--569.

\bibitem[Sabbadin et~al.(2014)Sabbadin, Ciancetta, and Moro]{Sabbadin2014JCIM}
Sabbadin, D., A.~Ciancetta, and S.~Moro, 2014.
\newblock {Perturbation of Fluid Dynamics Properties of Water Molecules during
  G Protein-Coupled Receptor--Ligand Recognition: The Human A2A Adenosine
  Receptor as a Key Study}.
\newblock \emph{J. Chem. Inf. Model.} 54:2846--2855.

\bibitem[Bai et~al.(2014)Bai, P{\'e}rez-S{\'a}nchez, Zhang, Shao, Shi, Liu, and
  Yao]{bai2014PCCP}
Bai, Q., H.~P{\'e}rez-S{\'a}nchez, Y.~Zhang, Y.~Shao, D.~Shi, H.~Liu, and
  X.~Yao, 2014.
\newblock {Ligand induced change of $\beta$2 adrenergic receptor from active to
  inactive conformation and its implication for the closed/open state of the
  water channel: insight from molecular dynamics simulation, free energy
  calculation and Markov state model analysis}.
\newblock \emph{Phys. Chem. Chem. Phys.} 16:15874--15885.

\bibitem[Grossfield et~al.(2008)Grossfield, Pitman, Feller, Soubias, and
  Gawrisch]{grossfield2008JMB}
Grossfield, A., M.~C. Pitman, S.~E. Feller, O.~Soubias, and K.~Gawrisch, 2008.
\newblock {Internal hydration increases during activation of the
  G-protein-coupled receptor rhodopsin}.
\newblock \emph{J. Mol. Biol.} 381:478--486.

\bibitem[Jard{\'o}n-Valadez et~al.(2010)Jard{\'o}n-Valadez, Bondar, and
  Tobias]{Jardon2010BJ}
Jard{\'o}n-Valadez, E., A.-N. Bondar, and D.~J. Tobias, 2010.
\newblock Coupling of retinal, protein, and water dynamics in squid rhodopsin.
\newblock \emph{Biophys. J.} 99:2200--2207.

\bibitem[Selent et~al.(2010)Selent, Sanz, Pastor, and {De
  Fabritils}]{Selent2010PLoSCompBiol}
Selent, J., F.~Sanz, M.~Pastor, and G.~{De Fabritils}, 2010.
\newblock {Induced Effects of Sodium Ions on Dopaminergic G-Protein Coupled
  Receptors}.
\newblock \emph{PLoS Comp. Biol.} 6:e1000884.

\bibitem[Liu et~al.(2012)Liu, Chun, Thompson, Chubukov, Xu, Katritch, Han,
  Roth, Heitman, IJzerman, Cherezov, and Stevens]{liu2012Science}
Liu, W., E.~Chun, A.~Thompson, P.~Chubukov, F.~Xu, V.~Katritch, G.~Han,
  C.~Roth, L.~Heitman, A.~IJzerman, V.~Cherezov, and R.~Stevens, 2012.
\newblock {Structural basis for allosteric regulation of GPCRs by sodium ions}.
\newblock \emph{Science} 337:232--236.

\bibitem[Venkatakrishnan et~al.(2013)Venkatakrishnan, Deupi, Lebon, Tate,
  Schertler, and Babu]{Venkatakrishnan2013Nature}
Venkatakrishnan, A., X.~Deupi, G.~Lebon, C.~G. Tate, G.~F. Schertler, and M.~M.
  Babu, 2013.
\newblock {Molecular signatures of G-protein-coupled receptors}.
\newblock \emph{Nature} 494:185--194.

\bibitem[Raghavender et~al.(2009)Raghavender, Aravinda, Shamala, Rai, and
  Balaram]{Raghavender2009JACS}
Raghavender, U.~S., S.~Aravinda, N.~Shamala, R.~Rai, and P.~Balaram, 2009.
\newblock Characterization of water wires inside hydrophobic tubular peptide
  structures.
\newblock \emph{J. Am. Chem. Soc.} 131:15130--15132.

\bibitem[Reddy et~al.(2010)Reddy, Straub, and Thirumalai]{Reddy2010PNAS}
Reddy, G., J.~E. Straub, and D.~Thirumalai, 2010.
\newblock Dry amyloid fibril assembly in a yeast prion peptide is mediated by
  long-lived structures containing water wires.
\newblock \emph{Proc. Natl. Acad. Sci. U. S. A.} 107:21459--21464.

\bibitem[Thirumalai et~al.(2011)Thirumalai, Reddy, and
  Straub]{thirumalai2011ACR}
Thirumalai, D., G.~Reddy, and J.~E. Straub, 2011.
\newblock Role of water in protein aggregation and amyloid polymorphism.
\newblock \emph{Acc. Chem. Res.} 45:83--92.

\bibitem[Xu et~al.(2011)Xu, Wu, Katritch, Han, Jacobson, Gao, Cherezov, and
  Stevens]{xu2011Science}
Xu, F., H.~Wu, V.~Katritch, G.~Han, K.~Jacobson, Z.~Gao, V.~Cherezov, and
  R.~Stevens, 2011.
\newblock {Structure of an agonist-bound human A2A adenosine receptor}.
\newblock \emph{Science} 332:322.

\bibitem[Jaakola et~al.(2008)Jaakola, Griffith, Hanson, Cherezov, Chien, Lane,
  IJzerman, and Stevens]{jaakola2008Science}
Jaakola, V., M.~Griffith, M.~Hanson, V.~Cherezov, E.~Chien, J.~Lane,
  A.~IJzerman, and R.~Stevens, 2008.
\newblock {The 2.6 angstrom crystal structure of a human A2A adenosine receptor
  bound to an antagonist}.
\newblock \emph{Science} 322:1211--1217.

\bibitem[Dor{\'e} et~al.(2011)Dor{\'e}, Robertson, Errey, Ng, Hollenstein,
  Tehan, Hurrell, Bennett, Congreve, Magnani, Tate, Weir, and
  Marshall]{dore2011structure}
Dor{\'e}, A.~S., N.~Robertson, J.~C. Errey, I.~Ng, K.~Hollenstein, B.~Tehan,
  E.~Hurrell, K.~Bennett, M.~Congreve, F.~Magnani, C.~G. Tate, M.~Weir, and
  F.~H. Marshall, 2011.
\newblock {Structure of the adenosine A$_{2A}$ receptor in complex with
  ZM241385 and the xanthines XAC and caffeine}.
\newblock \emph{Structure} 19:1283--1293.

\bibitem[Phillips et~al.(2005)Phillips, Braun, Wang, Gumbart, Tajkhorshid,
  Villa, Chipot, Skeel, Kale, and Schulten]{NAMD}
Phillips, J.~C., R.~Braun, W.~Wang, J.~Gumbart, E.~Tajkhorshid, E.~Villa,
  C.~Chipot, R.~D. Skeel, L.~Kale, and K.~Schulten, 2005.
\newblock {Scalable molecular dynamics with NAMD}.
\newblock \emph{J. Comp. Chem.} 26:1781--1802.

\bibitem[Buck et~al.(2006)Buck, Bouguet-Bonnet, Pastor, and
  MacKerell]{buck2006BJ}
Buck, M., S.~Bouguet-Bonnet, R.~W. Pastor, and A.~D. MacKerell, 2006.
\newblock {Importance of the CMAP correction to the CHARMM22 protein force
  field: dynamics of hen lysozyme}.
\newblock \emph{Biophys. J.} 90:L36--L38.

\bibitem[Zoete et~al.(2011)Zoete, Cuendet, Grosdidier, and
  Michielin]{Zoete2011JComputChem}
Zoete, V., M.~A. Cuendet, A.~Grosdidier, and O.~Michielin, 2011.
\newblock SwissParam: a fast force field generation tool for small organic
  molecules.
\newblock \emph{J. Comput. Chem.} 32:2359--2368.

\bibitem[Ku{\v{c}}erka et~al.(2011)Ku{\v{c}}erka, Nieh, and
  Katsaras]{kuvcerka2011BBA}
Ku{\v{c}}erka, N., M.-P. Nieh, and J.~Katsaras, 2011.
\newblock Fluid phase lipid areas and bilayer thicknesses of commonly used
  phosphatidylcholines as a function of temperature.
\newblock \emph{Biochimica et Biophysica Acta (BBA)-Biomembranes}
  1808:2761--2771.

\bibitem[Tsai et~al.(2013)Tsai, Lee, Huang, and Juwita]{tsai2013IJMS}
Tsai, H.-H.~G., J.-B. Lee, J.-M. Huang, and R.~Juwita, 2013.
\newblock A Molecular dynamics study of the structural and dynamical properties
  of putative arsenic substituted lipid bilayers.
\newblock \emph{Int. J. Mol. Sci.} 14:7702--7715.

\bibitem[Luzar and Chandler(1996)]{Luzar1996Nature}
Luzar, A., and D.~Chandler, 1996.
\newblock Hydrogen-bond kinetics in liquid water.
\newblock \emph{Nature} 379:55--57.

\bibitem[Yoon et~al.(2014)Yoon, Lin, Hyeon, and Thirumalai]{yoon2014JPCB}
Yoon, J., J.-C. Lin, C.~Hyeon, and D.~Thirumalai, 2014.
\newblock {Dynamical Transition and Heterogeneous Hydration Dynamics in RNA}.
\newblock \emph{J. Phys. Chem. B} 118:7910--7919.

\bibitem[Freeman(1979)]{freeman1979SocialNetworks}
Freeman, L., 1979.
\newblock Centrality in social networks conceptual clarification.
\newblock \emph{Social networks} 1:215--239.

\bibitem[Strogatz(2001)]{strogatz2001Nature}
Strogatz, S., 2001.
\newblock Exploring complex networks.
\newblock \emph{Nature} 410:268--276.

\bibitem[Albert et~al.(2000)Albert, Jeong, and Barab{\'a}si]{albert2000Nature}
Albert, R., H.~Jeong, and A.~Barab{\'a}si, 2000.
\newblock Error and attack tolerance of complex networks.
\newblock \emph{Nature} 406:378--382.

\bibitem[Newman(2005)]{newman2005SocialNetworks}
Newman, M., 2005.
\newblock A measure of betweenness centrality based on random walks.
\newblock \emph{Social networks} 27:39--54.

\bibitem[Greene and Higman(2003)]{greene2003JMB}
Greene, L., and V.~Higman, 2003.
\newblock Uncovering network systems within protein structures.
\newblock \emph{J. Mol. Biol.} 334:781--791.

\bibitem[Amitai et~al.(2004)Amitai, Shemesh, Sitbon, Shklar, Netanely, Venger,
  and Pietrokovski]{amitai2004JMB}
Amitai, G., A.~Shemesh, E.~Sitbon, M.~Shklar, D.~Netanely, I.~Venger, and
  S.~Pietrokovski, 2004.
\newblock Network analysis of protein structures identifies functional
  residues.
\newblock \emph{J. Mol. Biol.} 344:1135--1146.

\bibitem[Del~Sol et~al.(2006)Del~Sol, Fujihashi, Amoros, and
  Nussinov]{delSol2006MSB}
Del~Sol, A., H.~Fujihashi, D.~Amoros, and R.~Nussinov, 2006.
\newblock Residues crucial for maintaining short paths in network communication
  mediate signaling in proteins.
\newblock \emph{Mol. Sys. Biol.} 2:2006.0019.

\bibitem[Bagler and Sinha(2007)]{bagler2007Bioinfomatics}
Bagler, G., and S.~Sinha, 2007.
\newblock Assortative mixing in Protein Contact Networks and protein folding
  kinetics.
\newblock \emph{Bioinformatics} 23:1760--1767.

\bibitem[Vendruscolo et~al.(2002)Vendruscolo, Dokholyan, Paci, and
  Karplus]{vendruscolo2002PRE}
Vendruscolo, M., N.~Dokholyan, E.~Paci, and M.~Karplus, 2002.
\newblock Small-world view of the amino acids that play a key role in protein
  folding.
\newblock \emph{Phys. Rev. E.} 65:061910.

\bibitem[Dokholyan et~al.(2002)Dokholyan, Li, Ding, and
  Shakhnovich]{dokholyan2002PNAS}
Dokholyan, N., L.~Li, F.~Ding, and E.~Shakhnovich, 2002.
\newblock Topological determinants of protein folding.
\newblock \emph{Proc. Natl. Acad. Sci. U. S. A.} 99:8637.

\bibitem[Brandes(2001)]{brandes2001JMS}
Brandes, U., 2001.
\newblock A faster algorithm for betweenness centrality.
\newblock \emph{J. Math. Soc.} 25:163--177.

\bibitem[Franck et~al.(2014)Franck, Sokolovski, Kessler, Matalon,
  Gordon-Grossman, Han, Goldfarb, and Horovitz]{franck2014JACS}
Franck, J.~M., M.~Sokolovski, N.~Kessler, E.~Matalon, M.~Gordon-Grossman, S.-i.
  Han, D.~Goldfarb, and A.~Horovitz, 2014.
\newblock Probing water density and dynamics in the chaperonin GroEL cavity.
\newblock \emph{J. Am. Chem. Soc.} 136:9396--9403.

\bibitem[Franck et~al.(2015)Franck, Ding, Stone, Qin, and Han]{franck2015JACS}
Franck, J.~M., Y.~Ding, K.~Stone, P.~Z. Qin, and S.~Han, 2015.
\newblock Anomalously rapid hydration water diffusion dynamics near DNA
  surfaces.
\newblock \emph{J. Am. Chem. Soc.} 137:12013--12023.

\bibitem[Smart et~al.(1996)Smart, Neduvelil, Wang, Wallace, and
  Sansom]{Smart1996JMG}
Smart, O.~S., J.~G. Neduvelil, X.~Wang, B.~Wallace, and M.~S. Sansom, 1996.
\newblock HOLE: a program for the analysis of the pore dimensions of ion
  channel structural models.
\newblock \emph{J. Mol. Graph.} 14:354--360.

\bibitem[Li et~al.(2013)Li, Jonsson, Beuming, Shelley, and Voth]{Voth2013JACS}
Li, J., A.~L. Jonsson, T.~Beuming, J.~C. Shelley, and G.~A. Voth, 2013.
\newblock {Ligand-dependent activation and deactivation of the human adenosine
  A2A receptor}.
\newblock \emph{J. Am. Chem. Soc.} 135:8749--8759.

\bibitem[Aryal et~al.(2014)Aryal, Abd-Wahab, Bucci, Sansom, and
  Tucker]{aryal2014NatComm}
Aryal, P., F.~Abd-Wahab, G.~Bucci, M.~S. Sansom, and S.~J. Tucker, 2014.
\newblock {A hydrophobic barrier deep within the inner pore of the TWIK-1 K2P
  potassium channel}.
\newblock \emph{Nature communications} 5:4377.

\bibitem[Anishkin and Sukharev(2004)]{Anishkin2004BJ}
Anishkin, A., and S.~Sukharev, 2004.
\newblock Water dynamics and dewetting transitions in the small
  mechanosensitive channel MscS.
\newblock \emph{Biophys. J.} 86:2883--2895.

\bibitem[Isom and Dohlman(2015)]{Isom2015PNAS}
Isom, D.~G., and H.~G. Dohlman, 2015.
\newblock {Buried ionizable networks are an ancient hallmark of G
  protein-coupled receptor activation}.
\newblock \emph{Proc. Natl. Acad. Sci. U. S. A.} 112:5702--5707.

\bibitem[Aryal et~al.(2015)Aryal, Sansom, and Tucker]{aryal2015JMB}
Aryal, P., M.~S. Sansom, and S.~J. Tucker, 2015.
\newblock Hydrophobic gating in ion channels.
\newblock \emph{J. Mol. Biol.} 427:121--130.

\bibitem[Powell et~al.(2011)Powell, Cleary, Davenport, Shea, and
  Siwy]{powell2011NatureNanotech}
Powell, M.~R., L.~Cleary, M.~Davenport, K.~J. Shea, and Z.~S. Siwy, 2011.
\newblock Electric-field-induced wetting and dewetting in single hydrophobic
  nanopores.
\newblock \emph{Nature Nanotechnology} 6:798--802.

\bibitem[Kobilka and Deupi(2007)]{kobilka2007TPS}
Kobilka, B.~K., and X.~Deupi, 2007.
\newblock {Conformational complexity of G-protein-coupled receptors}.
\newblock \emph{Trends Pharmaco. Sci.} 28:397--406.

\bibitem[J. et~al.(2012)J., Wrabl, and Motlagh]{Hilser12ARB}
J., H.~V., J.~O. Wrabl, and H.~N. Motlagh, 2012.
\newblock Structural and Energetic Basis of Allostery.
\newblock \emph{Annu. Rev. Biophys.} 41:585--609.

\bibitem[Motlagh et~al.(2014)Motlagh, Wrabl, Li, and Hilser]{Motlagh2014Nature}
Motlagh, H.~N., J.~O. Wrabl, J.~Li, and V.~J. Hilser, 2014.
\newblock The ensemble nature of allostery.
\newblock \emph{Nature} 508:331--339.

\bibitem[Lockless and Ranganathan(1999)]{Lockless99Science}
Lockless, S.~W., and R.~Ranganathan, 1999.
\newblock Evolutionarily Conserved Pathways of Energetic Connectivity in
  Protein Families.
\newblock \emph{Science} 286:295--299.

\bibitem[Halabi et~al.(2009)Halabi, Rivoire, Leibler, and
  Ranganathan]{halabi2009cell}
Halabi, N., O.~Rivoire, S.~Leibler, and R.~Ranganathan, 2009.
\newblock Protein sectors: evolutionary units of three-dimensional structure.
\newblock \emph{Cell} 138:774--786.

\bibitem[Zheng et~al.(2005)Zheng, Brooks, Doniach, and
  Thirumalai]{Zheng05Structure}
Zheng, W., B.~R. Brooks, S.~Doniach, and D.~Thirumalai, 2005.
\newblock Network of Dynamically Important Residues in the Open/Closed
  Transition in Polymerases Is Strongly Conserved.
\newblock \emph{Structure} 13:565--577.

\bibitem[Ribeiro and Ortiz(2014)]{ribeiro2014JCTC}
Ribeiro, A.~A., and V.~Ortiz, 2014.
\newblock Determination of signaling pathways in proteins through network
  theory: importance of the topology.
\newblock \emph{J. Chem. Theor. Comp.} 10:1762--1769.

\bibitem[Feher et~al.(2014)Feher, Durrant, Van~Wart, and Amaro]{feher2014COSB}
Feher, V.~A., J.~D. Durrant, A.~T. Van~Wart, and R.~E. Amaro, 2014.
\newblock Computational approaches to mapping allosteric pathways.
\newblock \emph{Curr. Opin. Struct. Biol.} 25:98--103.

\bibitem[Di~Paola and Giuliani(2015)]{diPaola2015COSB}
Di~Paola, L., and A.~Giuliani, 2015.
\newblock Protein contact network topology: a natural language for allostery.
\newblock \emph{Curr. Opin. Struct. Biol.} 31:43--48.

\bibitem[Leitner(2016)]{leitner2016JPCB}
Leitner, D.~M., 2016.
\newblock {Water-Mediated Energy Dynamics in a Homodimeric Hemoglobin}.
\newblock \emph{J. Phys. Chem. B} 120:4019--4027.

\bibitem[Yoon et~al.(2013)Yoon, Thirumalai, and Hyeon]{Yoon13JACS}
Yoon, J., D.~Thirumalai, and C.~Hyeon, 2013.
\newblock {Urea-induced denaturation of preQ1-riboswitch}.
\newblock \emph{J. Am. Chem. Soc.} 135:12112--12121.

\bibitem[R{\'e}at et~al.(2000)R{\'e}at, Dunn, Ferrand, Finney, Daniel, and
  Smith]{Reat2000PNAS}
R{\'e}at, V., R.~Dunn, M.~Ferrand, J.~L. Finney, R.~M. Daniel, and J.~C. Smith,
  2000.
\newblock Solvent dependence of dynamic transitions in protein solutions.
\newblock \emph{Proc. Natl. Acad. Sci. U. S. A.} 97:9961--9966.

\bibitem[Fitter et~al.(1996)Fitter, Lechner, Buldt, and
  Dencher]{Fitter1996PNAS}
Fitter, J., R.~Lechner, G.~Buldt, and N.~Dencher, 1996.
\newblock Internal molecular motions of bacteriorhodopsin: hydration-induced
  flexibility studied by quasielastic incoherent neutron scattering using
  oriented purple membranes.
\newblock \emph{Proc. Natl. Acad. Sci. U. S. A.} 93:7600--7605.

\bibitem[Li et~al.(2007)Li, Hassanali, Kao, Zhong, and Singer]{li2007JACS}
Li, T., A.~A. Hassanali, Y.-T. Kao, D.~Zhong, and S.~J. Singer, 2007.
\newblock Hydration dynamics and time scales of coupled water-protein
  fluctuations.
\newblock \emph{J. Am. Chem. Soc.} 129:3376--3382.

\bibitem[Wood et~al.(2007)Wood, Plazanet, Gabel, Kessler, Oesterhelt, Tobias,
  Zaccai, and Weik]{wood2007PNAS}
Wood, K., M.~Plazanet, F.~Gabel, B.~Kessler, D.~Oesterhelt, D.~Tobias,
  G.~Zaccai, and M.~Weik, 2007.
\newblock Coupling of protein and hydration-water dynamics in biological
  membranes.
\newblock \emph{Proc. Natl. Acad. Sci. U. S. A.} 104:18049--18054.

\bibitem[Fogarty and Laage(2014)]{fogarty2014JPCB}
Fogarty, A.~C., and D.~Laage, 2014.
\newblock Water dynamics in protein hydration shells: the molecular origins of
  the dynamical perturbation.
\newblock \emph{J. Phys. Chem. B.} 118:7715--7729.

\end{thebibliography}

\end{document}